\documentclass[12pt]{article}

\global\arraycolsep=1pt

\usepackage{amsbsy,amssymb,latexsym,amsfonts, amsmath}
\usepackage{mathrsfs}
\usepackage{graphicx}

\usepackage{amsbsy,amssymb,latexsym,amsfonts,amsmath}
\usepackage{mathrsfs}
\usepackage{graphicx}
\usepackage{bm}
\usepackage{here}
\usepackage{cite}

\numberwithin{equation}{section}
\newcommand{\bel}[1]{\begin{equation}\label{#1}}                     
\newcommand{\bal}[1]{\begin{eqnarray}\label{#1}}                     
\newcommand{\bea}{\begin{eqnarray}}
\newcommand{\eea}{\end{eqnarray}}

\newcommand{\im}{\mathrm{i}}
\newcommand{\ex}{\mathrm{e}}
\newcommand{\de}{\mathrm{d}}
\newcommand{\dis}{\displaystyle}

\newcommand{\qq}{\qquad}
\newcommand{\mat}[1]{\begin{pmatrix} #1 \end{pmatrix}}
\renewcommand{\thefootnote}{\fnsymbol{footnote}}

\newcommand{\be}{\begin{equation}}
\newcommand{\ee}{\end{equation}}
\newcommand{\beq}{\begin{equation}}
\newcommand{\eeq}{\end{equation}}

\def\CD{{\mathcal D}}

\def\CA{{\mathcal A}}

\def\CN{\mathcal N}

\newcommand{\CF}{{\mathcal F}}
\newcommand{\CW}{{\mathcal W}}

\begin{document}
%
%
\begin{titlepage}
\begin{flushright}
\normalsize
~~~~
OCU-PHYS 427\\
July, 2015 \\
\end{flushright}

\vspace{15pt}

\begin{center}
{\LARGE
Developments of theory of effective prepotential} \\
\vspace{5pt}
{\LARGE  from extended Seiberg-Witten system }\\
\vspace{5pt}  
 {\LARGE and matrix models }\\
\end{center}

\vspace{23pt}

\begin{center}
{ H. Itoyama$^{a, b}$\footnote{e-mail: itoyama@sci.osaka-cu.ac.jp} 
and
R. Yoshioka$^b$\footnote{e-mail: yoshioka@sci.osaka-cu.ac.jp}
}\\
%
\vspace{18pt}
%

$^a$ \it Department of Mathematics and Physics, Graduate School of Science\\
Osaka City University\\
\vspace{5pt}

$^b$ \it Osaka City University Advanced Mathematical Institute (OCAMI)

\vspace{5pt}

3-3-138, Sugimoto, Sumiyoshi-ku, Osaka, 558-8585, Japan \\

\end{center}
%
\vspace{20pt}
\begin{center}
Abstract\\
\end{center}

This is a semi-pedagogical review of a medium size on the exact determination of and the role played by the
   low energy effective prepotential ${\cal F}$ in QFT with (broken) extended supersymmetry,
    which began with the work of Seiberg and Witten in 1994.
   While paying an attention to an overall view of this subject lasting long over the two decades,
    we probe several corners marked in  the three major stages of the developments, emphasizing
    uses of the deformation theory on the attendant Riemann surface as well as its close relation
    to matrix models. Examples picked here in different contexts tell us that
    the effective prepotential is to be identified as 
    the suitably defined free energy $F$ of a matrix model:
    ${\cal F} = F$.

 To be submitted to PTEP as an invited review article and based in part on the talk delivered by one of 
  the authors (H.I.) in the workshop held 
  at Shizuoka University, Shizuoka, Japan, on December 5, 2014.


\vfill

\setcounter{footnote}{0}
\renewcommand{\thefootnote}{\arabic{footnote}}

\end{titlepage}

\renewcommand{\thefootnote}{\arabic{footnote}}
\setcounter{footnote}{0}


\section{Introduction} \label{intro}

  The notion of effective action plays a vital role in the modern treatment 
  of quantum field theory. (See, for instance, \cite{WK1974T,PS11995A}.)
In this review article, we deal with a special class of  low energy effective actions that
 are controlled by (broken) extended rigid supersymmetry in four spacetime dimensions and
 permit exact determination exploiting integrals on
 a Riemann surface in question. A main object in such study is  the low energy effective
  prepotential to be denoted by $\mathcal{F}$ generically in this paper, which has proven to be central
  not only in the original case of unbroken ${\cal N}=2$ supersymmetry initiated by the work of
  Seiberg-Witten \cite{SW9407E,SW9408M} but also
   in  the case where this symmetry is broken 
   by the vacuum or  by the superpotential.
   The review will be presented basically in a chronological order, following 
   the three major stages of the developments that took place during the periods $1994 \sim$, $2002 \sim$ and $2009 \sim$.
   Each of the three subsequent sections will explain  pieces of work done in its respective period.
   
   An emphasis will be put on the deformation theory
   of the effective prepotential on the Riemann surface as an extension of the Seiberg-Witten
    system consisting of the curve, the meromorphic differential and the period
      as well as its close relation to matrix models.
    
 We conclude from the examples taken here in the different contexts that
  the effective prepotential is in fact identified as the suitably defined 
  free energy $F$ of a matrix model: ${\cal F} = F$. 
  While this is hardly a surprising conclusion from the point of view of mathematics 
  of integrable systems and soliton hierarchies, 
  the number of examples in QFT where this is explicitly materialized is not 
  large enough. 
  This note may serve to improve the situation.

  In the next section, after presenting the curve for $\mathcal{N}=2$, $SU(N)$ pure super Yang-Mills theory
   as a spectral curve of the periodic Toda chain, we discuss the deformation of the effective prepotential
  by placing higher order poles to the original meromorphic differential. We give a derivation of the formula
   which the meromorphic differential extended this way obeys.

  In section three, we discuss the degeneration phenomenon of the Riemann surface necessary to describe
   the $\mathcal{N}=1$ vacua that lie in the confining phase and introduce
the prepotential having gluino condensates as variables. 
We apply the formalism in section 2 here, 
 and describe the situation by the use of mixed second 
 derivatives. 
After discussing the emergence of the matrix model curve 
 and giving sample calculation, 
 we finish the section with the case of 
 spontaneously broken $\mathcal{N}=2$ supersymmetry 
 in order to illustrate the role played by 
 the two distinct singlet operators 
 one of which is the QFT counterpart of the matrix model resolvent.
 
In section four, 
 we go back to the situation of $\mathcal{N}=2$ and
 discuss the developments associated with the AGT relation 
 and the upgraded treatment of the all-genus instanton 
 partition function and 
 therefore the deformation of the Seiberg-Witten curve to 
 its noncommutative counterpart.
A finite $N$ and $\beta$-deformed matrix model 
 with filling fractions specified 
 emerge as an integral representation 
 of the conformal/W block and 
 we discuss the direct evaluation of its $q$-expansion as 
 the Selberg integral. 
We finish the section with mentioning some of 
 the more recent developments.

Please note that the model or theory hops from one to the other 
 as the sections proceed and that each section has 
 its open ending, indicating calls for further developments 
 of this long lasting subject.

\section{effective prepotential from extended Seiberg-Witten system} 

We will not give here an account of the construction of 
 the curve itself \cite{SW9407E,SW9408M,KLYT9411S,AF9411T,
 HO9505O,APS9505T,AS9509T}, 
(for a recent review, for instance, \cite{TT1108S})
nor its connection to classical integrable system
\cite{Kric1977M,Kric1977I,OP1981C,GKMM9505I,MW9509I,NT9509W,DW9510S,EY9510P,
 Mart9510I,GM9510T,MW9511I,IM19511I,GHL9512S,IM29512P,
 IM9601I,LO9605D}.
Also omitted is the discussion associated 
 with the WDVV equation, for which we direct the readers
 to \cite{Witt1991T,DVV1991T,MMM9607W}
 as well as references contained in 
 \cite{IM40211T,Mars1999S}. 

\subsection{curves, periods and meromorphic differentials} 
The list of papers which discuss subjects closely related to 
 that of this subsection include 
\cite{SW9407E,SW9408M,KLYT9411S,AF9411T,
GKMM9505I,HO9505O,APS9505T,Mato9506I,MW9509I,NT9509W,
AS9509T,DW9510S,EY9510P,Mart9510I,GM9510T,MW9511I,IM19511I,
GHL9512S,IM29512P,IM9601I,AN9603I,GMMM9603N,GMMM9604A,
BM9605N,SW9607G,GHL9608K,DKP9609T,Nekr9609F,BMT9610S,
Dona9705S,Klem9705O,DP9709C,GGM9710S,GH9711F,MM97115,
KO9801P,DP9804S,BMMM9812S,BMMM9902T,FGNR9906D,BMMM9906O,
MM9912C,HM9912C,Mars1999S,BM0009S,GM0011I,TT1108S}.

Let us recall the most typical situation
 and consider the low energy effective action (LEEA) 
 for $\mathcal{N}=2$, $SU(N)$ pure super Yang-Mills theory. 
The symmetry of LEEA at the scale much smaller than that of the W boson mass 
 is $U(1)^{N-1}$. 
The  relevant curve 
 is a hyperelliptic Riemann surface of genus $N-1$ described as
\begin{align}
Y^2 &= P_N^2(x) - 4 \Lambda^{2N}, \\ 
{\rm where} \nonumber \\
P_N(x) &= \langle \det (x {\bf 1} - \Phi) \rangle  
\equiv \prod_{i=1}^N (x - {p_i}) = x^N - \sum_{k=2}^N {u_k} x^{N-k} \nonumber\\
&= \sum_{k=0}^N s_k(h_{\ell}) x^{N-k}. 
\end{align}
Here, 
\begin{align}
{h_{\ell}} = \frac{1}{\ell} \langle {\rm tr} \Phi^{\ell} \rangle 
= \frac{1}{\ell} \sum_{i=1}^N {p_i^{\ell}}, 
~~~~~~ \ell = 2,3,\cdots,N,
\end{align}
 and  $s_k(h_{\ell})$ are the appropriate Schur polynomials.
 Introducing the spectral parameter $z$, 
 we write the curve as that of the periodic Toda chain:
\begin{align}
P_N(x) &= z + \frac{{\Lambda^{2N}}}{z},  \\
Y &= z - \frac{{\Lambda^{2N}}}{z}. 
\end{align}
\begin{figure}[h]
\centering
  \includegraphics[height=3cm]{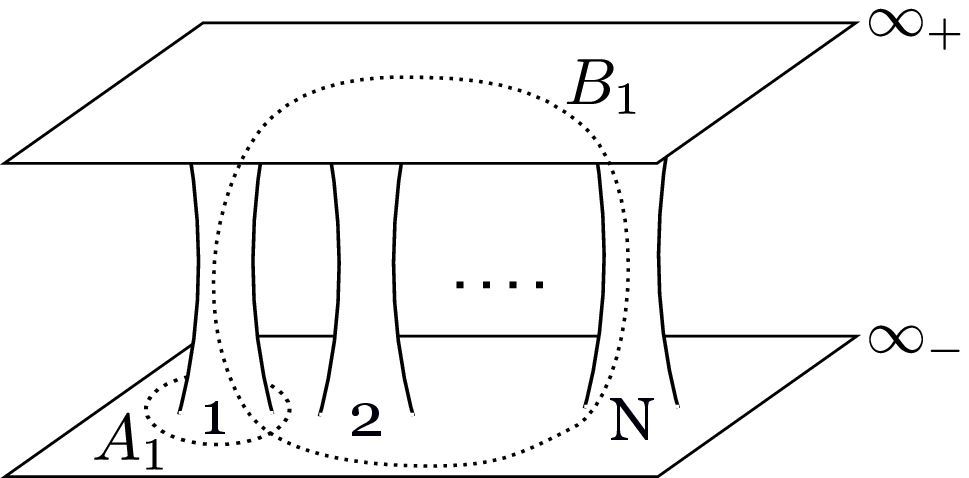}
\caption{
\label{fig1}
}
\end{figure}%

\noindent
The distinguished meromorphic differential  for the construction of the effective prepotential  is given by
\begin{align*}
\de \hat{S}_{\rm SW} = x \de \log z = x {t(x)} \de x,~~~
{t(x)} = \frac{P'_N}{\sqrt{P_N^2 - 4 \Lambda^2}}. 
\end{align*}

The characteristic feature of this is the existence of double poles at
 $\infty_{\pm}$.
  Later in this section, we interpret this to be the case where only $T_1$ has been turned on.

\noindent
The defining property is that the moduli derivatives are holomorphic:  
\begin{align}
\left. \frac{\partial}{\partial u_k} \de \hat{S}_{\rm SW} \right|_{{z},\Lambda} 
&= \frac{x^{N-k}}{Y} \de x, \label{ModDer1}\\
\text{or} \hspace{25mm} & \notag \\
\left. \frac{\partial}{\partial u_k} \de \hat{S}_{\rm SW} \right|_{{x},\Lambda} 
&= \frac{x^{N-k}}{Y} \de x  - \de\left( \frac{x^{N-k+1}}{Y} \right).
\label{ModDer2} 
\end{align}
The prepotential $\mathcal{F}_{\rm SW}$ is introduced implicitly by 
 the A cycle and B cycle integrations on the Riemann surface:
\beq
 a_i = \oint_{A_i} \de \hat{S}_{\rm SW}, ~~~~~
 \frac{\partial \mathcal{F}_{\rm SW}}{\partial a_i} (=a^D_i)
  = \oint_{B_i} \de \hat{S}_{\rm  SW}. 
\eeq
While $u_k$ possess invariant meaning both in the moduli space of the 
 Riemann surface and in the integrable system, 
 it is these constant background fields or Coulomb moduli $a_i$, 
 $a^D_i = \frac{\partial \mathcal{F}_{\rm SW}}{\partial a_i}$ 
 which are directly related to the observables through the BPS formula. 
 The moduli derivatives are coordinate dependent as we see
 in eqs. \eqref{ModDer1} and \eqref{ModDer2}. 
The final expression for $\mathcal{F}_{\rm SW}$ is going to be 
 coordinate independent. 
This is supported by the pieces of evidence
we present here that the effective prepotential is identified as the 
 free energy of a matrix model.

\subsection{Whitham deformation of the prepotential and the appearance of ``thermodynamic" relation}
The list of papers which discuss subjects closely related to 
 that of this subsection include 
\cite{Kric9205T,Dubr1992H,AK9505T,IM29512P,
CVV9710I,KY9712R,Carr9712R,GMMM9802R,EMM9805W,Moro9903W,Taka9905W,EGMM9911N,
MZ9912W,Mars1999S,GM0011I}.

 We would now like to  review the deformation of the  effective prepotential above
  which we have denoted by $\mathcal{F}_{\rm SW}$.
  The basic idea of this extended theory of effective prepotential often referred to as Whitham deformation is 
  to deform both moduli of the Riemann surface and the meromorphic differential above consistently 
 without losing the defining properties:
\be\label{(bullet)}
\de \hat{S}_{\rm SW} \to {\de \hat{S}};~~~ 
~~ \frac{\partial}{\partial h_k} {\de \hat{S}}|_{{*},\Lambda}
 = \text{holomorphic}. 
\ee
We have adopted the choice that  $z$ is fixed when the moduli derivatives are taken. 
 We carry out the deformation  by adding higher order poles to the original meromorphic differential
  containing the double poles. Let us denote the local coordinates in their neighborhood   generically by $\xi$   and 
\be
\label{higherorderpole}
{\xi = z^{\mp \frac{1}{N}}~~ {\rm or}~~ x^{-1}}\;.
\ee
  In order to describe the deformation, let us introduce  a set of meromorphic differentials $\de \Omega_{\ell}$
   that satisfy
\be
 \de \Omega_{\ell} = \xi^{-\ell-1} \de \xi
+ \text{non-singular \; part}~~~~ \ell =1.2,3, \cdots \;\;. 
\ee
  We are still left with the ambiguities that  any linear combination of the canonical
   holomorphic differentials $\de \omega_i$ can be added to the right hand side.
In order to remove these,  let us require a set of conditions
\be \label{(*)}
 \oint_{{A_i}} \de \Omega_{\ell} = 0 \;\;.
\ee
The ones which are not subject to the conditions eq.\eqref{(*)} are denoted by $\de \widehat{\Omega}_{\ell}$.

 Let us first state the formula 
\be\label{(star)}
 {\de \hat{S} 
 = \sum_{i=1}^g {a^i} \de \omega_i + \sum_{\ell\geq 1} {T_{\ell}} \de \Omega_{\ell}} \;\;
\ee
and outline its derivation below. 
 As before, $a^i$ are defined to be the local coordinates in the moduli space 
 \be
{a^i} \equiv \oint_{A_i} \de \hat{S} \;\;,
\ee
while $T_{\ell}$, referred to as time variables or T moduli, are given by 
 \be
{T_{\ell}}= \underset{\xi = 0}{\rm res} \xi^{\ell} \de\hat{S} \;\;,
\ee
once eq.\eqref{(star)} is established. One then regards
$a^i$ and $T_{\ell}$ as independent, taking $h_k$  dependent: ${h_k = h_k}({a^i, T_{\ell}})$.
The (extended) effective prepotential ${\mathcal{F}}({a^i,T_{\ell}})$ is introduced via
\be\label{(**)}
 \frac{\partial {\mathcal{F}}}{\partial {a^i}} 
 = \oint_{B^i} \de \hat{S}, ~~~
 \frac{\partial {\mathcal{F}}}{\partial {T_{\ell}}} 
 = \frac{1}{2 \pi \im \ell}{\rm res} \xi^{-\ell}  \de \hat{S} 
 \equiv {\mathcal{H}_{\ell+1}}{(h_k)}\;. 
\ee

The derivation of  eq.\eqref{(star)}  begins with
the introduction of the time variables $T_{\ell}$ via 
 a solution $\de \hat{S}(T_{\ell}|h)$ to eq.\eqref{(bullet)}, 
\be\label{(1)}
 \text{namely,}~~~~
\frac{\partial \de \hat{S}}{\partial T_{\ell}} = \de \Omega_{\ell}\;\;,
 ~~~~~ \text{and~~ hence} 
~~~~~~~~ \frac{\partial a_i}{\partial T_{\ell}} = 0\;. 
\ee
 In terms of our intermediate bases $\de \widehat{\Omega}_{\ell}$, eq.\eqref{(bullet)} reads
\begin{align}
 & \frac{\partial}{\partial h_k} \de \hat{\Omega}_{\ell} 
 = \sum_{i=1}^g \sigma_{ki}^{(\ell)} \de \omega_i, \\
\text{ while}~~~~~~ &\de \widehat{\Omega}_{\ell} 
 = \de {\Omega}_{\ell} + \sum_{i=1}^{g} c_i^{(\ell)} \de \omega_i \;\;, 
 \label{(2)}
\end{align}
as the difference between $\de \hat{\Omega}_{\ell}$ and $\de {\Omega}_{\ell}$ can be spanned by
 the holomorphic differentials.
Expand the solutions as
\begin{align}
&\de \hat{S} = \sum_m \beta_m(T) \de \widehat{\Omega}_m(h) \;\;, 
 \label{(+)} \\
\text{hence}~~~~~~~~~~
&\frac{\partial \de\hat{S}}{\partial T_n} 
= \sum_m \left( \frac{\partial \beta_m}{\partial T_n} 
\de\widehat{\Omega}_m + \beta_m \sum_k \frac{\partial h_k}{\partial T_n} 
\sum_{i=1}^g \sigma_{ki}^{(m)} \de \omega_i \right). 
\label{(3)}
\end{align}
Exploiting  eq.\eqref{(1)}, eq.\eqref{(2)} and eq.\eqref{(3)}, we obtain
\begin{align}
& \frac{\partial \beta_m}{\partial T_n} = \delta_{m,n} 
 ~~~~~ {\rm i.e.}~~\beta_m(T) = T_m
 \label{(4)}\\
 \text{as~well~as}~~~~~~  & \sum_k \frac{\partial h_k}{\partial T_n} \left( \sum_m T_m \sigma_{ki}^{(m)} \right) 
 = - c_i^{(n)}.    
 \label{(5)}
\end{align}
 Substituting  eq.\eqref{(4)} and eq.\eqref{(2)}  into eq.\eqref{(+)}, we obtain 
\be 
 \de\hat{S} = \sum_m T_m \de\Omega_m + \sum_m T_m \sum_i c_i^{(m)} \de \omega_i, 
\ee 
 whose integrations over the $A_i$ cycles yield
\be
 a_i = \sum_m T_m c_i^{(m)}. 
\ee
This shows eq.\eqref{(star)}.

\subsection{connection with the planar free energy of matrix models}

Already at this stage of the developments, a keen connection of the extended Seiberg-Witten system
 with the construction of matrix models in general, or more specifically, the similarity of 
 the effective prepotentials with the (planar) free energy of matrix models was visible.
In fact, starting from the homogeneity of the moduli and 
 the prepotential, it is possible to derive an integral expression
  for $\CF$ which resembles that of matrix model planar 
  free energy in terms of the density one-form on the eigenvalue coordinate. 
See, eq. (4.12) of \cite{IM29512P}. Also \cite{GKMM9505I,NT9509W,EY9510P}. 

One of the goals of the present review is to put together subsequent several developments 
 that took place and have made this phenomenon more prominent. 
These are presented in the next two sections.

\section{Gluino condensate prepotential} \label{II}
One major use of the deformation theory of the effective prepotential  presented above
 took place in the context of gluino condensate prepotential built 
  on various $\mathcal{N}=1$ vacua in contrast to $\CF_{\rm SW}$ 
  and its extension in section 2.
We first consider the case in which the  breaking to $\mathcal{N}=1$ from $\mathcal{N}=2$ supersymmetry
 is caused by the superpotential in the action. Later we will contrast this with the case in which 
 $\mathcal{N}=2$ is broken spontaneously to $\mathcal{N}=1$ at the tree level
 \cite{APT9512S,AT9604D,FIS0409S,FIS0410U,FIS0503P}.
 \footnote{Actually, supersymmetry is broken dynamically in the metastable 
 vacua in both cases as was demonstrated in \cite{IM1109D,IM1301D}
  in the Hartree-Fock approximation.}
\subsection{degeneration phenomenon and mixed second derivatives}
The list of papers which discuss subjects closely related to 
 that of this subsection include 
\cite{IM1301D,IM1109D,VY1982A,TVY1983S,NSVZ1983I,Koni1984A,Seib1988S,NSVZ1989N,ILS9403E,DS9503D,EFGI9603M,
TY9706A,GVW9906C,CIV0103A,CV0206N,DV0206M,DV0207O,DV0208A,CM0209M,
Ferr0210O,Ferr0211Q,IM40211T,IM50211E,IM60212C,Mato0212S,
CMMV0301D,IM70301G,BBDW0304N,Holl0305C,BBDW0305F,AMM0310P,IK0312W,GMT0601O}.

Let's fix  an action to work with: it is a $U(N)$ gauge theory  consisting of 
  adjoint vector superfields and chiral superfields with canonical kinematic
  factors and the superpotential turned on in the $\mathcal{N}=2$ action  drives the system to its $\mathcal{N}=1$ vacua.
 
 As  a phenomenon occurring on a Riemann surface, we consider the situation  
  where a degeneration  takes place and
  some of the cycles coalesce to form a new set of cycles.  
  As for the description of the low energy effective action (LEEA),  
  some of the original Coulomb moduli disappear and the product of these $U(1)$ s
  gets replaced by non-Abelian gauge symmetry ${\dis{\prod_{i=1}^n SU(N_i)}}$.
  We tabulate  these pictures  below.
\begin{table}[H]
\begin{center}
\begin{tabular}{lccc}
\hline
&$\mathcal{N}=2$ & & $\mathcal{N}=1$  \\ \hline \hline
&$U(N)$ pure SYM & $\xrightarrow[\substack{{\rm deformed} \\ \text{by superpotential}}]{}$ 
& $\dis{\prod_{i=1}^n U(N_i)}$   \\\\
&\multicolumn{3}{c}{$\dis{g_{n+1} \int \de^2\theta {\rm tr} W_{n+1}(\Phi)}$ 
~~~ such~that $\dis{W_{n+1}'(x) = \prod_{i=1}^n(x - \alpha_i)}$} \\\\
LEEA &$\underset{\lower2ex \vbox to-2pt{\hbox{\small{\rm Coulomb}}}}
{\dis{U(1)^{N-1} \times U(1)}}$ 
 & $\longrightarrow$ 
& $\underset{\lower2ex \vbox to -2pt{\hbox{\small{\rm Coulomb}}}}
 {\dis{U(1)^{n-1} \times U(1)}}
 \times 
 \underset{\lower0ex \vbox to 0pt{\hbox{\small{\rm confining}}}}
 {\dis{\prod_{i=1}^n SU(N_i)}}$\\\\
RS & 
\raisebox{-9mm}{\includegraphics[height=20mm]{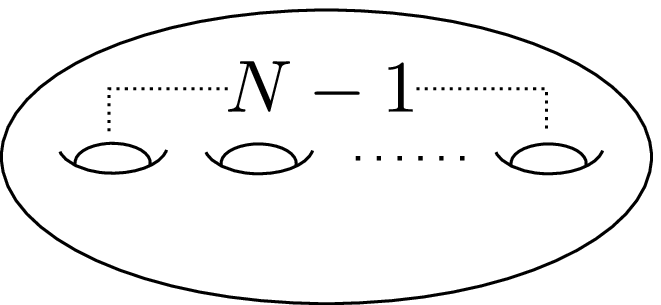}}~~
&$\xrightarrow[{\rm degeneration}]{}$&\raisebox{-9mm}{\includegraphics[height=20mm]{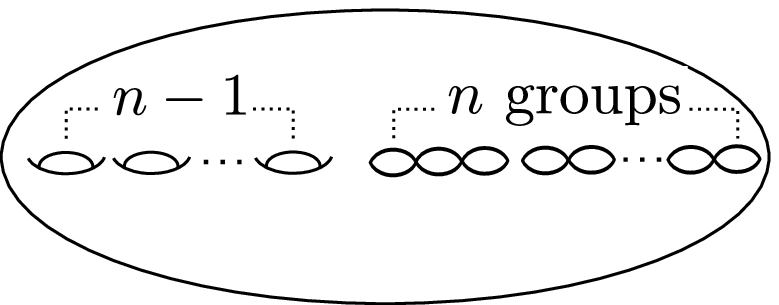}}
\end{tabular}
\end{center}
\end{table}
The $\mathcal{N}=1$ vacua are
 labelled by the set of order parameters representing gluino condensates:
\be
 S_i \propto {\rm Tr}_{SU(N_i)} W^{\alpha} W_{\alpha}, ~~~~~
 i=1, \cdots, n.
\ee
 The proportionality constant will be fixed in  subsequent subsections.

We now review, following the observation made in \cite{IK0312W} 
 that the condition for a curve to degenerate or
 factorize is given by that the kernel of the matrix made of the mixed second derivatives of
 the deformed prepotential be nontrivial.

Continuing with the general discussion of subsection 2.2, 
 let us first note  that  we obtain  two different expressions for the mixed second derivatives from eq.\eqref{(**)}:
\be
 \frac{\partial^2 \mathcal{F}}{\partial a^i \partial T_{\ell}} 
 = \oint_{B_i} \de \Omega_{\ell} = \frac{1}{2\pi \im \ell} {\rm res} \xi^{-\ell} \de \omega_i,
 ~~~~~i=1,\cdots,N-1, ~~~~ \ell: \text{ positive integers}
\ee
We impose the condition
\be\label{(***)}
 {\rm ker}\frac{\partial^2 \mathcal{F}}{\partial a^i \partial T_{\ell}} \neq 0, 
 ~~~ {\rm or}~~~ 
 {\rm rank} \frac{\partial^2 \mathcal{F}}{\partial a^i \partial T_{\ell}} \leq N-2. 
\ee 

\noindent
Eq.\eqref{(***)} has following straightforward implications:  \\
i) there  exists
a nonvanishing column vector   
$\mat{c^1, ~ c^2, ~ \cdots ~ c^{N-1}, ~ \cdots}^t$ such that
\be
\label{cycledegeneration}
0 = \sum_{\ell} \frac{\partial^2 \mathcal{F}}{\partial a^i \partial T_{\ell}} c^{\ell} 
 \underset{eq. \eqref{(1)}}{=} \sum_{\ell} \oint_{B_i} \de \Omega_{\ell} c^{\ell} 
 \underset{eq. \eqref{higherorderpole}}{=} \frac{1}{2\pi \im} \underset{\xi = 0}{\rm res} 
   \left( \sum_{\ell} \frac{c^{\ell}}{\ell} \xi^{-\ell} \right)_{+} \de \omega_i \;\;.
\ee
Here, we have exploited eq. \eqref{(1)} in the second equality 
 and eq. \eqref{higherorderpole} in the third equality.
 The former equality implies that
$\displaystyle{
\de \widetilde{\Omega} \equiv \sum_{\ell} c^{\ell} \de \Omega_{\ell}}
$
has vanishing periods over all $A_i$ \&  $B^i$ cycles.Then one can integrate this form along any path
 ending with a point $z$ to define a function holomorphic except at punctures.  
 As for the order of the poles at the punctures, 
 it is generically arbitrary according to the construction.
 But  this is contradictory to the Weierstrass gap theorem \footnote{ 
  The Weierstrass gap theorem states that
\begin{align}
 {\rm for~ a~  given~ Riemann~ surface}~ M,~ {\rm with~  genus}~ g, {\rm and~ a~ point~} P \in M,   \\
 {\rm and~} g~{\rm integers~ satisfying} ~~1 = n_1 < n_2 < \cdots < n_g < 2g,
\end{align}
there does NOT exist a function $f$ holomorphic on $M\backslash \{P\}$ 
 with a pole of order $n_j$ at $P$.  }
 derived from the Riemann-Roch theorem.
To avoid  a contradiction,  we must have a degeneration.

ii) there exists
 a nonvanishing row vector 
 $\mat{\widetilde{c}_1, \widetilde{c}_2, \cdots, \widetilde{c}_{N-1} }$ such that 
\be
\label{Hdegeneration}
0 = \sum_{i=1}^{N-1} \widetilde{c}_i 
\frac{\partial^2 \mathcal{F}}{\partial a^i \partial T_{\ell}} 
= \sum_i \widetilde{c}_i \frac{\partial \mathcal{H}_{\ell+1}}{\partial a^i} 
\ee
 in accordance with  the second formula of eq. \eqref{(**)}. 
 Eq. \eqref{Hdegeneration}  follows from
\be
\label{hdegeneration}
\sum_i \widetilde{c}_i \frac{\partial h_{\ell + 1}}{\partial a^i} = 0\;,
\ee
  which is regarded as the statement of the vanishing discriminant.
The moduli depend actually on less than $N-1$ arguments.

\subsection{emergence of the matrix model curve}

The list of papers which discuss subjects closely related to 
 that of this subsection include 
 \cite{DHKS0209E,DHKS0209M,FO0210C,Bere0210Q,DHK0210S,
Gors0210K,ACFH0210E,McGr0211A,BR0211E,AKMV0211M,Gopa0211N,Feng0211S,KMT0211G,
KM0211C,NSW0211M,IM40211T,IM70301G,AM0309T,AFH0311A,IK0312W}.

 Once we are convinced of the degeneration of the surface, we can proceed further by 
  factorizing the original curve, which, in the current example,
  is the hyperelliptic one.

Let $n-1$ be the number of genus
after the degeneration.  Following \cite{CIV0103A,CV0206N}, we state
\be
\left( \begin{array}{l}
 \displaystyle{Y^2 = H_{N-1}(x)^2 F_{2n}(x)}, \\\\
 P_N'(x) = H_{N-n}(x) R_{n-1}(x) \;.
 \end{array}
\right.
\ee

Finally let us examine the last equality of eq.\eqref{cycledegeneration}. Let 
\be
 \sum_{\ell=1}^{N-1} \frac{c_{\ell}}{\ell} \left( \xi^{-\ell} \right)_+ 
 \equiv {W'_{k+1}(x)} {\equiv \prod_{j=1}^k(x-\alpha_j)}. \\
\ee
and $\frac{x^{j-1}}{\sqrt{F_{2n}}}$ serve as bases of  the holomorphic differentials of the reduced Riemann surface.
 Actually, only the $j= 1 \sim n-1$ differentials are holomorphic and  the $j=n$ one has been added
 through the blow-up process, which physically implies  that the overall  $U(1)$ fails to decouple.
We obtain 
\begin{align}
\label{gettingmmcurve}
0 = \underset{x = \infty}{\rm res} 
 \left( W'_{k+1}(x) \frac{x^{j-1}}{\sqrt{F_{2n}}} \right), \\
{\rm and~ therefore~~~}
\frac{W'_{k+1}}{\sqrt{F_{2n}}} = Q_{k-n}(x) 
 + \sum_{\ell > n} \frac{\beta_{\ell}}{x^{\ell}} .
\end{align}

\be
 {\rm We ~obtain~~~} y^2 \equiv F_{2n} Q_{k-n}^2 = W'_{k+1} + f_{k-1}. 
\ee
 Here, $f_{k-1}$ is a polynomial of degree $k-1$.
This is the curve appearing in the $k$-cut solution of the matrix model.

 We still need to see  that $W_{k+1}(x)$ introduced above is in fact a tree level superpotential.
This is easily done by taking the classical limit $\Lambda = 0$:
\be
 Y = z = \prod_{\ell=1}^N (x - p_{\ell}). 
\ee
The original Seiberg-Witten differential becomes
\begin{align}
\de \hat{S}_{\rm SW}^{\rm (class)} &= x \sum_{i=1}^N \frac{1}{x-p_i}\de x \;,\\
{\rm which ~ is~ equal~ to~~~}   &= \sum_{i=1}^N p_i \de \omega_i + N \de x.
\end{align}
 Here, we have used  that the canonical holomorphic differential becomes
\be
 \de \omega_i^{\rm (class)} = \frac{\de x}{x - p_i} \;\;
\ee
in this limit.
The period integrals over the $A_i$ cycles just pick up the residues at the poles $p_i$:
\be
{a_i^{\rm (class)}} = p_i. 
\ee
The degeneration in this limit is described as
\be
 z = \prod_{j=1}^n (x -\beta_j)^{N_j},~~ \sum_{j=1}^n N_j = N\;.
\ee
In fact,  the $N_j$ poles coalesce at $\beta_j, ~ j=1,\cdots, n\;$ and   the canonical
holomorphic differentials on the degenerate curve are
\be
\de \omega_j^{\rm (class, red)} = \frac{\de x}{x - \beta_j} \;\;.
\ee
The condition eq. \eqref{gettingmmcurve} becomes
\be
 0 = \underset{x=0}{\rm res} \left(
 W'_{k+1}(x) \de \omega_j^{\rm (class,red)}\right), ~~~~~ 
 j= 1, \cdots, n, \;\;\;,
\ee
 which tells us that ${\beta_j}$ must coincide with one of the roots ${\alpha_j}$ 
 of $W'_{k+1}$. 
The vev's of the adjoint scalar fields are thus constrained to the extrema of $W_{k+1}$.

Let us set $k=n$ for simplicity.
We have the reduced curve of $g = n-1$:
\be
\label{reducedcurve}
 y^2 = W'_{n+1}(x;\alpha_j)^2 + f_{n-1}(x). 
\ee
  and  let us denote the coefficients of the polynomial $f_{k-1}$ by $b_{\ell}(\alpha_j)$,
 temporarily forgetting the $\alpha_j$ dependence.
We also mention here that the full set of parameters (moduli) of the model 
realized by the curve eq. \eqref{reducedcurve} is $2n$ dimensional 
 and  can be  represented by the cut lengths and cut positions:
\be
 {\rm dim(moduli)} 
 = 2n \approx \text{cut lengths} + \text{cut positions}. 
\ee
\begin{figure}[H]
\centering
  \includegraphics[height=3.5cm]{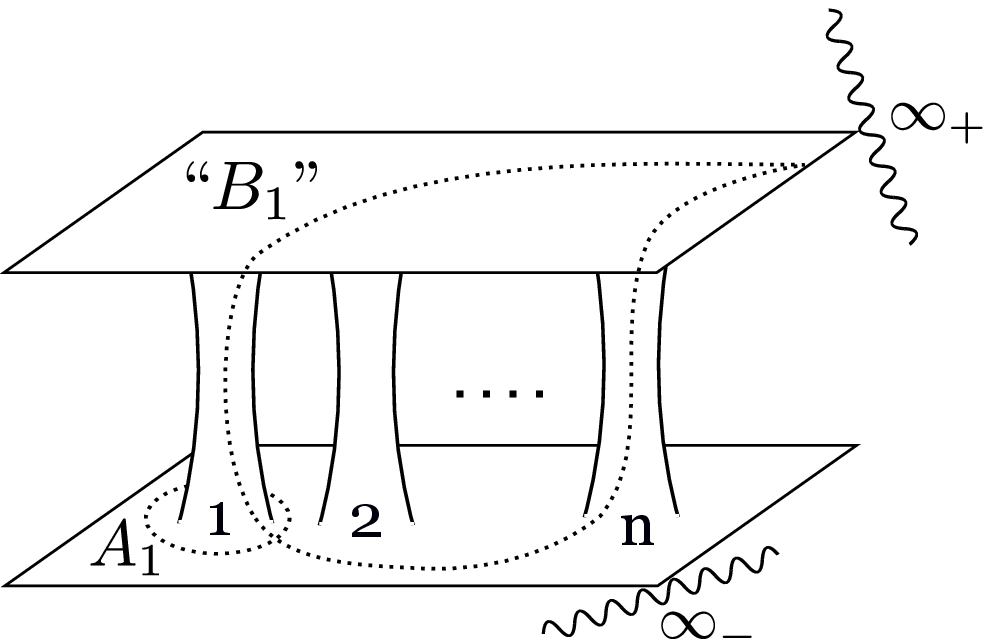}
\caption{
\label{fig2}
}
\end{figure}%

\subsection{practical calculation}
The list of papers which discuss subjects closely related to 
 that of this subsection include 
\cite{DGKV0210P,DGLV0211P,Suzu0211P,DJ0211E,NSW0211T,IM60212C,
IM70301G,DP0301O,KKM0303D,Gome0405E,Aoya0504T,HMPS0804N}. 

 Let us now proceed to discuss the use of this machinery in calculation.
  As the condensates $S_i$  are quantum mechanical in nature, 
  one can develop loop
   expansion using these, including 
   the Veneziano-Yankielowicz term
   which contains the logarithmic singularity \cite{VY1982A}.
 The first question to be raised is
  what the distinguished meromorphic differential is to be used for such calculation.
 It must be "almost" holomorphic after the $b_{\ell}$ derivatives are taken. 
 Recall that the bases of the "holomorphic" differentials are taken as $\frac{x^{j-1}}{y}, ~ j=1,\cdots,n-1,n$.  
Rather obviously,  such differential is found as
\be
\label{matdiff}
 \de \hat{S}_{\rm mat} = y(x) \de x,~~~~~~
  {\rm with}~~~~~~~~~~~~~T_2, \cdots,T_n,~T_{n+1}~~~~~~~~{\rm turned~~on}. 
\ee
 As before, the effective prepotential is introduced through the period integrals
\be \label{(star2)}
\left({
\begin{array}{llc}
&~~~~~\dis{{S_i} = \oint_{{A_i}} \de \hat{S}_{\rm mat}~~  i = 1,\cdots,n}, ~~~~~~~~~~
&
\\
{\rm and} &&\\
&~~~~~\displaystyle{\frac{\partial {\mathcal{F}}}{\partial {S_i}}} 
= 2 \int_{i~{\rm edge}}^{\rm cutoff} {\de \hat{S}_{\rm mat}} \;.
&
\end{array}}
\right.
\ee 
 We have, however, no reason to set
\be
\dis{S \equiv \sum_{i=1}^n S_i}
= \int_{\prod_{i=1}^{n} ~ \!\!\!\!^{\cup}A_i} \de \hat{S}_{\rm mat} 
\ee
equal to zero. This tells us the presence of the cutoff at the infinities of the surface.

The expansion of $\mathcal{F}$ in $S_i$ was done in \cite{IM60212C},  
 exploiting eq. \eqref{(star2)} 
 and the small cut expansion as an intermediate step originally. 
This provided 
the answer  given below for $\mathcal{F}$ to 
 the cubic order  in $S_i$ (eq. \eqref{F3} - \eqref{Delta}). 
Yet, there exists a simpler procedure, namely, a calculus from $T$ moduli 
 thanks to the machinery discussed in the present review. The $T$ moduli are easily identified  as
 \be
 {T_{m+1}} = \underset{\infty}{\rm res} ~x^{-m-1} \de \hat{S}_{\rm mat} = g u_m,~~~~~
u_m = (-)^{n-m} e_{n-m}^{(\alpha)}, 
\ee
where
\be
 e_m(\alpha)= \sum_{i_1 < \cdots < i_m} \alpha_{i_1}\cdots \alpha_{i_m} \;.
\ee
The dependence of  the prepotential on the $T$ moduli is determined by  the equations 
\be\label{(starstar)}
{{{\frac{1}{g}\frac{\partial \mathcal{F}}{\partial u_{\ell}} 
= {\frac{\partial \mathcal{F}}{\partial T_{\ell+1}}} 
= \frac{1}{\ell+1} \underset{\infty}{\rm res} 
(x^{\ell+1} - \Lambda^{\ell+1}) \de \hat{S}_{\rm mat}}}}.  
\ee
Here $\Lambda^{\ell+1}$ is the term introduced in \cite{IM70301G} 
 in order to match with the computation done earlier. 
 In order to carry out this task,  we introduce intermediate expansion variables $\tilde{S}_i$
and parameterize   the matrix model curve eq. \eqref{reducedcurve} by
\be
 f_{n-1}(x) = \sum_{i=1}^n {\tilde{S}_i} \prod_{j(\neq i)}^n (x - \alpha_j) 
 = W'_{n+1}(x) \sum_{i=1}^n \frac{{\tilde{S}_i}}{x - \alpha_i}. 
\ee
The differential $\de \hat{S}_{\rm mat}$ of eq. \eqref{matdiff} has a 
 straightforward expansion in $\tilde{S}_i$. 
Therefore, $A_i$ cycle integrations followed 
 by the inversion provide an expansion of
 $\tilde{S}_i$  in $S_j$
\be
 \tilde{S}_i = S_i + \frac{1}{2g} 
 \sum_{j,k} \frac{1}{\alpha_{ij} \alpha_{ik} \Delta_i} S_j S_k + \cdots \;\;.
\ee
Here, we have introduced
$\alpha_{ij} = \alpha_{i} - \alpha_{j}$, and $ \displaystyle {\Delta_i = \prod_{j \neq i} \alpha_{ij} } $.
 Another useful machinery is the $T_m$ moduli derivatives of the roots $\alpha_i$ of the
  superpotential, which read
$\dis{\frac{\partial \alpha_j}{\partial u_m} 
 = - \frac{\alpha_j^m}{\Delta_j}}$.  Using these, the right hand side of eq. \eqref{(starstar)}
 is evaluated as
\begin{align} 
  \sum_i S_i \left( \frac{\partial W_{n+1}(\alpha_i)}{\partial u_{\ell}} 
 - \frac{\partial W_{n+1} (\Lambda)}{\partial u_{\ell}} \right) 
 - \frac{1}{4} \sum_{j<k} \left( S_j^2 + S_k^2 - 4S_jS_k \right) 
 \frac{\partial}{\partial u_{\ell}} \log \alpha_{jk} + \cdots \;,
\end{align}
  which is trivially integrated in $u_m$ to provide  an answer.
Let us mention that this procedure is straightforwardly 
 generalizable to higher order contributions
 in $S_i$ and that
the terms independent of $\alpha_i$ can be easily obtained
by several other methods.

The expansion form of $\mathcal{F}(S|\alpha)$ 
 which we managed to have proposed in \cite{IM60212C} is 
\footnote{In transition to this equation, there is a 
 change in the normalization, which we avoid discussing here. 
 See \cite{IM70301G}}
\begin{align}
2\pi i {{\cal F}(S|\alpha)} = 4\pi ig_{n+1}\left(W_{n+1}({\Lambda})\sum_i S_i 
- \sum_i W_{n+1}({\alpha_i})S_i\right) -
(\sum_{i} S_i)^2\log{\Lambda} +
\notag\\ 
+ \frac{1}{2} \sum_{i=1}^n S_i^2\left(\log \frac{S_i}{4}\ - \frac{3}{2}\right) 
- \frac{1}{2}\sum_{i<j}^n(S_i^2 - 4S_iS_j + S_j^2)\log {\alpha_{ij}} + 
\sum_{k=1}^\infty \frac{{{\cal F}_{k+2}(S|\alpha)}}{(i\pi g_{n+1})^k}. 
\end{align}

Here, we have denoted by
$\mathcal{F}_{k+2}(S|\alpha)$  the contributions of the
$k+2$ order polynomials in $S_i$. The explicit answer for $\mathcal{F}_{3}(S|\alpha)$ is
\begin{align} 
&{\cal F}_3(S|\alpha) = \sum_{i=1}^n u_i(\alpha)S_i^3
+ \sum_{i\neq j}^nu_{i;j}(\alpha) S_i^2S_j +
\sum_{i<j<k}^n u_{ijk}(\alpha)S_iS_jS_k,  
 \label{F3}\\
&u_i(\alpha) = \frac{1}{6}
\left(
-\sum_{j(\neq i)}\frac{1}{\alpha_{ij}^2\Delta_j} +
\frac{1}{4\Delta_i}\sum_{\stackrel{j<k}{j,k(\neq i)}}
\frac{1}{\alpha_{ij}\alpha_{ik}}
\right),  \label{ui}\\
&u_{i;j}(\alpha) = \frac{1}{4}
\left(
-\frac{3}{\alpha_{ij}^2\Delta_i}
+ \frac{2}{\alpha_{ij}^2\Delta_j}
- \frac{2}{\alpha_{ij}\Delta_i}\sum_{k\neq i,j}\frac{1}{\alpha_{ik}}
\right), \label{uij}\\
&u_{ijk}(\alpha) = \frac{1}{\alpha_{ij}\alpha_{ik}\Delta_i} +
 \frac{1}{\alpha_{ji}\alpha_{jk}\Delta_j} +
 \frac{1}{\alpha_{ki}\alpha_{kj}\Delta_k},  \label{uij}\\
&\Delta_i = W_{n+1}^{\prime\prime}(\alpha_i) = \prod_{j\neq i}^n\alpha_{ij}.
\label{Delta} 
\end{align}

For the computation of higher orders as well as the inclusion of matter, 
 see, for instance, \cite{NSW0211T,Gome0405E,Aoya0504T,HMPS0804N}. 

\subsection{case of spontaneously broken $\mathcal{N}=2$ supersymmetry and Konishi anomaly equation}

The list of papers which discuss subjects closely related to 
 that of this subsection include 
 \cite{
CGP1986C,CGP1986A,HP1986P,HLP1986S,
FGP9510M,APT9512S,FGP9512S,ADKM9604S,AT9604D,Porr9609S,PP9702P,IZ9710M,
GVW9906C,TV9912R,
CDSW0211C,Tach0211D,CSW0301P,IKan0304S,DGN0304K,ACDG0305G,
FIS0409S,KP0409O,FIS0410U,FIS0503P,FIS0510P,Merl0511F,
Ferr0602T,FIS0602S,IMaru0603U,Fuji0609P,FIS0611S,MN0612E,
Ferr0701T,IMaru0704D,Ferr0709E,IMS0709N,Maru0710E,FW0710C,IMaru0710D,
Sait0711T,Ferr0804T,ABSV0804E,OT0806E,NNT0806I,
Maru0904Q,IMM0909L,FI2009S}.

 The $\mathcal{N} = 2$ effective action is completely characterized by the effective prepotential
  while, in the $\mathcal{N} = 1$ case,  a typical observable is (the matter induced part of) the
   effective superpotential. The interplay of these two upon the degeneration of the original Riemann surface is
most clearly seen  by dealing with 
the case of spontaneously broken $\mathcal{N}=2$ supersymmetry. This case accomplishes a continuous deformation
from one to the other by tuning the electric and magnetic Fayet-Iliopoulos parameters.
The action $S^{\mathcal{F}_{\rm in}}_{\mathcal{N} = 2}$ realizing this is given by
\begin{align}
\label{FIS}
    S_{\CN=2}^{\CF_{\rm in}}   
    =&    \int d^4 x d^4 \theta 
           \left[
         - \frac{i}{2} {\rm Tr} 
           \left(  \bar{\Phi} e^{ad V} 
           \frac{\partial \CF_{\rm in}(\Phi)}{\partial \Phi}
         - h.c.
           \right)
         + {\xi} V^0 
           \right] 
           \notag \\   
	&   + \left[
           \int d^4 x d^2 \theta
           \left(
         - \frac{i}{4} 
           \frac{\partial^2 \CF_{\rm in}(\Phi)}{\partial \Phi^a \partial \Phi^b}
           \CW^{\alpha a} \CW^b_{\alpha}
         + {e} \Phi^0
         + {m} \frac{\partial \CF_{\rm in}(\Phi)}{\partial \Phi^0}
           \right)
         + h.c.
           \right].
\end{align}
 Here, 
$\xi, e, m$ are the electric and magnetic F-I terms and  
 we vary these to interpolate the two ends, keeping
$\tilde{g}_{\ell} = m g_{\ell}~ (\ell \geq 2)$ fixed:   
\begin{table}[H]
\begin{center}
\begin{tabular}{ccccc}
\hline 
&large $(\xi,e,m)$ & & small $(\xi,e,m)$ \\ \hline \hline \\
$S_{\CN=1}$
& $\CN=1$ 
& $\xleftarrow[\hspace{3cm}]
{\displaystyle{S_{\mathcal{N}=2}^{\mathcal{F_{\rm in}}}}}$ 
& $\CN=2$  & $S_{\CN=2}$
\end{tabular}
\end{center}
\end{table}

In this subsection, we have denoted by the symbol $\mathcal{F}_{\rm in}$ 
 an input function in the effective action eq. \eqref{FIS}.  
For definiteness, we let the
 function $\CF_{\rm in}$   be a single trace function of a polynomial in $\Phi$
\be
 \CF_{\rm in}(\Phi) = \sum_{\ell=1}^{n+1} \frac{g_{\ell}}{(\ell+1)!} {\rm Tr} \Phi^{\ell+1},
  ~~~ {\rm deg} \CF_{\rm in} = n+2,
\ee
and the matter induced part of the effective superpotential $W_{\rm eff}$
be
\be
 \ex^{i \int d^4 x (d^2 \theta W_{\rm eff} + h.c. + (\text{D-term}))}
     =     \int \CD \Phi \CD \bar{\Phi} \ex^{i S_{\CN=2}^{\CF_{\rm in}}}  .
\ee

Let us now turn to the
 generalized Konishi anomaly equation.
It is the anomalous Ward identity of the theory given by
 eq.\eqref{FIS} and is derived by considering
 a response of the system under for the general local transformation 
 $\delta \Phi = f(\Phi, \mathcal{W})$:  
\begin{align}
- \left< \frac{1}{64 \pi^2} 
           \left[ 
           \CW^\alpha , 
           \left[ \CW_\alpha , \frac{\partial f}{\partial \Phi_{ij}}
           \right]
           \right]_{ij}
           \right>   
     =     \left<
           {\rm Tr} f W'(\Phi)
           \right>
         - \left<
           \frac{i}{4} {\rm Tr} (f \CF_{\rm in}'''(\Phi) \CW^\alpha \CW_\alpha)
           \right>.
\end{align}
The left-hand side is the contribution of 
 the Konishi anomaly \cite{Koni1984A}, which  arises from the behavior
of the functional integral measure under the transformation 
 \cite{Fuji1979P,Fuji1980P}.
Introducing the two generating functions, 
 we recast this into the following set of equations
 \cite{IMaru0704D}: 
\begin{align}
R(z)
    & \equiv - \frac{1}{64 \pi^2} \left<
           {\rm Tr} 
           \frac{\CW^\alpha \CW_\alpha}{z - \Phi}
           \right>,
           \nonumber \\
    T(z)   
    & \equiv   \left< {\rm Tr} 
           \frac{1}{z - \Phi}
           \right>, \\
{R(z)^2}
    & {\hspace{1mm}=    W'(z) R(z) + \frac{1}{4} f(z)}, 
\label{3.46}\\
    {2 R(z) T(z)}
    & {\hspace{1mm}=    W'(z) T(z) + 16 \pi^2 i \CF_{\rm in}'''(z) R(z) + \frac{1}{4} c(z)},
\label{3.47}
\end{align}
where $f(z)$ and $c(z)$ are polynomials of degree $n-1$ and, 
 with some abuse in notation, 
\be \label{F'''}
 \CF_{\rm in}'''(z) 
     =     \sum_{\ell=2}^{n+1} \frac{g_\ell z^{\ell-2}}{(\ell - 2)!}
     =     \frac{W''(z)}{m}.
\ee
The explicit form of $f(z)$ and that of $c(z)$ are not really needed in what follows. 

Let us make a few comments on this set of equations.
The equation for $R(z)$ 
 is identical in form to that of the planar loop equation of the one-matrix model for the resolvent.  This fact is
shared by  the theory in the large FI term limit, 
 namely, ${\mathcal N}=1$ theory
  of adjoint vector superfields and chiral superfields with 
  a general superpotential \cite{CDSW0211C}.
The equation for $T(z)$, on the other hand, contains  the cubic derivatives 
 in ${\mathcal F}_{\rm in}$ and is distinct from that in the large FI term limit.
This, in fact, leads us to the deformation of the formula connecting the effective superpotential
 with the object identified as the matrix model free energy from its well-known expression 
 \cite{DV0206M,DV0207O,DV0208A} in $S_{\CN=1}$, namely, the one in the large FI term limit.

Our final goal in this subsection is  
 to derive a formula for the effective superpotential.  
Let us define the one point functions as
\be
v_\ell
     =   - \frac{1}{64 \pi^2} \langle {\rm Tr} \CW^\alpha \CW_\alpha \Phi^\ell \rangle,
           ~~~~
    u_\ell
     =     \langle {\rm Tr} \Phi^\ell \rangle,
           ~~~~
           {\rm for}~~ 1 \leq \ell \leq n+1.
\ee 
In terms of $v_{\ell}$ we define $F$ as
\be
 \frac{\partial F}{\partial g_\ell}
     =     \frac{m}{\ell!} v_\ell,
           ~~~~
           {\rm for}~~ 1 \leq \ell \leq n+1.
\ee
Using $F$,  we can state the relation to be proven:
\be 
W_{eff}
= \sum_i N_i \frac{\partial F}{\partial S_i}
 + \frac{16 \pi^2 i}{m} \sum_{\ell = 2}^{n+1} 
 g_\ell \frac{\partial F}{\partial g_{\ell - 1}}.
 \label{3.51}
\ee
Before proceeding to the proof of this relation, 
let us go back to eqs. \eqref{3.46} and \eqref{3.47} 
to obtain the complete information. 
We consider the most general case that the gauge symmetry 
 $U(N)$ is broken to ${\displaystyle \prod_{i=1}^{k}} U(N_i)$ with 
 $k<n,~ {\displaystyle \sum_{i=1}^k } N_i = N$. 
The indices $i, j, \cdots$ run from 1 to $k$ 
while the indices $I, J, \cdots$ run from 1 to $N$. 
Of course, $N_I =0$ $(I = k+1, \cdots, n)$. 
Solving eq. \eqref{3.46}, we obtain     
\bea
    R(z)
     =     \frac{1}{2} \left( W'(z) - \sqrt{W'(z)^2 + f(z)} \right), 
           \label{1}
\eea
where the Riemann surface $\Sigma$ is genus $n-1$ but 
 its $A_I$ cycles for $I = k+1, \cdots, n$ are vanishing. 
We conclude that the meromorphic function lives 
 on a factorized curve 
\bea
    y^2
     =     W'(z)^2 + f(z)
     =     N_{n-k}(z)^2 F_{2k}(z),
           \label{2}
\eea
\bea
    y^2_{red}
     =     F_{2k}(z).
           \label{reduce}
\eea
Here $N_{n-k}(z)$, $F_{2k}(z)$ are polynomials 
 of degree $n-k$ and $2k$ respectively. 
On the other hand, substituting eq. \eqref{1} into eq. \eqref{3.47}, 
we obtain 
\bea
    T(z)   
     =   - \frac{c(z)}{4 \sqrt{W'(z)^2 + f(z)}}
         + 8 \pi^2 i 
           \left(      
           \CF_{\rm in}'''(z) 
         - \frac{W'(z) \CF_{\rm in}'''(z)}{\sqrt{W'(z)^2 + f(z)}}     
           \right).
           \label{4}
\eea

Let us list a few formulas that are obtained from eq. \eqref{1} 
 directly. The first set is \cite{Gopa0211N}
\bea
    \frac{\partial R(z)}{\partial S_i}
     =  \frac{g_i(z)}{4 \sqrt{F_{2k}(z)}}, ~~~~~
     \frac{\partial f(z)}{\partial S_i} 
     = N_{n-k}(z) g_i(z), ~~~~~ i = 1, \cdots, k.           
      \label{3}
\eea
Here $\frac{g_i(z)}{4 \sqrt{F_{2k}(z)}}$, $i= 1, \cdots, n$ is 
 a set of normalized holomorphic functions, as  is easily seen by taking the derivatives of 
 the A cycle integrations. 
Also, define $h(z) = - {\displaystyle \sum_i} N_i g_i(z)$. 
The second one is 
\bea
    \frac{16 \pi^2 i}{m} 
    \sum_{\ell = 1}^n g_{\ell + 1} \frac{\partial R(z)}{\partial g_\ell}
    &=&    8 \pi^2 i
           \left(      
           \CF_{\rm in}'''(z) 
         - \frac{W'(z) \CF_{\rm in}'''(z)}{\sqrt{W'(z)^2 + f(z)}}     
           \right)
           \nonumber \\
         ~~~~~~~~~~~
     & &  + \frac{16 \pi^2 i}{m}
          \left(
           \frac{- {\displaystyle \sum_{\ell = 1}^n } g_{\ell+1} \partial f(z)/\partial g_\ell}{4 \sqrt{W'(z)^2 + f(z)}}
           \right),
           \label{5}
\eea
where we have used eq. \eqref{F'''}. 

The proof eq. \eqref{3.51} goes by observing that 
 it is equivalent to the truncation of the following equation 
 up to the first $n+1$ terms in the $1/z$ expansion, 
\bea
    T(z)
     =     \sum_i N_i \frac{\partial R(z)}{\partial S_i} 
         + \frac{16 \pi^2 i}{m} \sum_{\ell=2}^{n+1} g_{\ell} 
         \frac{\partial R(z)}{\partial g_{\ell-1}}.  
           \label{6}
\eea
Substituting eqs. \eqref{1}, \eqref{4}, \eqref{3} and \eqref{5}
 into eq. \eqref{6}, we see that the proof becomes complete 
 as  soon as we obtain 
\begin{align}
    D(z)
     &=    \frac{16 \pi^2 i}{m}
           \sum_{\ell = 1}^n g_{\ell+1} \frac{\partial f(z)}{\partial g_\ell},
           \label{7}
           \\ 
  {\rm where~~~}   D(z)
     &\equiv
           c(z) - N_{n-k} h(z). 
    \label{8}
\end{align}

Observe that there are two expressions for $N_i$: 
\bea
    N_i
     =     \oint_{A_i} T(z) \de z 
     = - \oint_{A_i} \frac{h(z)}{4 \sqrt{F_{2k}(z)}} \de z,
           ~~~~~
            \;\;\; i= 1,\ldots,k, 
           \label{9}
\eea 
and therefore 
\be
 \oint _{A_I} \left( T(z) + \frac{h(z)}{4 \sqrt{F_{2k}(z)}}
 \right) = 0, ~~~~~ I = 1, \cdots, n. 
 \label{10}
\ee
Another consistency condition is 
\be
 0 = \frac{\partial S_I}{\partial g_{\ell}}
 = \frac{\partial}{\partial g_{\ell}} \oint_{A_I} R(z), 
 ~~~~~ I = 1, \cdots, n.  
 \label{11} 
\ee
Eliminating $\CF'''(z) - \frac{W'(z) \CF'''(z)}
 {\sqrt{W'(z)^2 + f(z)}}$ 
  in the integrand of eq. \eqref{10} and
  that of eq. \eqref{11}, 
 we obtain
\bea
    0
     =     \oint_{A_I}
           \frac{D(z) - \frac{16 \pi^2 i }{m}\sum_{\ell = 1}^n g_{\ell+1} 
           \frac{\partial f(z)}{\partial g_\ell}}
           {4 \sqrt{W'(z)^2 + f(z)}} \de z.
    \label{12}
\eea
Expanding the integrand of this equation by a set of holomorphic 
 differentials $\frac{z^\ell \de z}{\sqrt{W'(z)^2 + f(z)}}, 
 ~ \ell = 0, \cdots, n-1$ of the original curve, we deduce eq. \eqref{7}.  

\section{AGT relation and 2d-4d connection via matrices} \label{III}

The contents of the two preceding sections later had 
the upgraded treatments mentioned in the introduction.   
In this section we 
 outline these developments triggered 
 by the work \cite{AGT0906L}. 

\subsection{Instanton partition function: What is $Z_{\rm inst}^{\epsilon_1,\epsilon_2}$? }
The list of papers which discuss subjects closely related to 
 that of this subsection include  
\cite{LMNS9509F,LNS9711I,Naka1999L,Nekr0206S,FP0208A,NY0306I,NO0306S,NY0311L,NY0505I}.

Let us recall  that the low energy effective action (LEEA) of 
 $\mathcal{N}=2$ $SU(N_c)$ SUSY gauge theory
 is specified by the effective prepotential denoted in this section by
 $\mathcal{F}_{\rm SW}(a_i)$ and that  it has undetermined VEV called Coulomb moduli
$a_i = \langle\phi_i\rangle$. The bare gauge coupling and the $\theta$ parameter are grouped into
\be
 q_{_{\text{bare}}} = e^{\pi i \tau_{_{\text{bare}}}}, ~~~~~
\tau_{_{\text{bare}}} = \frac{\theta}{\pi} + \frac{8 \pi i}{g^2_{_{\text{bare}}}}, 
\ee
 and $\mathcal{F}_{\rm SW} (a_i)$ consists of the one-loop contribution and the instanton sum
\be
 \mathcal{F}_{\rm SW} = 
 \mathcal{F}_{\text{1-loop}} + \mathcal{F}_{\text{inst}}^{(\rm SW)}, 
\ee

It was shown in \cite{Nekr0206S} that 
$\CF_{\rm inst}^{(\rm SW)}$ is microscopically calculable in the presence 
 of $\Omega$ background
equipped with the deformation parameters $\epsilon_1$ and $\epsilon_2$ as 
\be
 Z_{\text{inst}}( \epsilon_1,\epsilon_2,a_i;q) = 
\exp\left(\frac{1}{\epsilon_1\epsilon_2} 
\mathcal{F}_{\text{inst}} (\epsilon_1,\epsilon_2,a_i)\right), ~~~~ 
\mathcal{F}_{\text{inst}} (0,0,a_i) = \mathcal{F}_{\text{inst}}^{(\rm SW)}. 
\ee
 The corrections to the original $\mathcal{F}_{\text{inst}}^{(\rm SW)}$  are regarded as 
higher orders in  the genus expansion with $g_s^2 = -\epsilon_1\epsilon_2$. 
Its expansion in $q$ is 
computable by the localization technique with $\epsilon_1, \epsilon_2$ acting as Gaussian cutoffs. 

\be
 Z_{\text{inst}} (a_i, \epsilon_1,\epsilon_2,; q) 
 \equiv \sum_{k=0}^{\infty}  Z_k  q^k ,
\ee
  where
\be \label{vol:moduli}
 Z_k \equiv \int_{\widetilde{M}_k} {\bf 1}_{\epsilon_1,\epsilon_2, a_i}.
\ee
 is the ``volume" of the $k$-instanton moduli space.

Let $T^{N_c-1}$ be the maximal torus of the gauge group $SU(N_c)$.  
Since we also have the maximal torus $T^2$ of $SO(4)$, namely, 
the global symmetry of $\textbf{R}^4$,  
the $T = T^2 \times T^{N_c-1}$ action can 
 be defined on the instanton moduli space.  
Then the integral in eq. \eqref{vol:moduli} are computed $T$-equivariantly  
 and consequently we obtain the regularized results.  
According to the localization formula, 
  eq. \eqref{vol:moduli} is reduced to the summation of 
 the contribution from the fixed points 
 which are parametrized by $N_c$ Young diagrams 
 $\vec{Y} = (Y^{(1)}, \cdots, Y^{(N_c)})$, 
\be
 Z_k = \sum_{|\vec{Y}| = k} Z_{\vec{Y}},  
\ee
where $|\vec{Y}| = \sum_{i=i}^{N_c} |Y^{(i)}|$ is the total number of boxes. 
Each $Z_{\vec{Y}}$ is provided through a combinatorial method. 

\subsection{$\beta$-ensemble of quiver matrix model and noncommutative curve}
The list of papers which discuss subjects closely related to 
 that of this subsection include 
\cite{BIPZ1978P,BK1990E,DS1990S,GM1990N,GM1990A,MM1990O,AJM1990M,Davi1990L,
Harr1991L,deBo1991M,IMat1991N,IMat1991W,MMM1991G,I9111M, 
AIMZ9112S,AS9112O,KMMM9208C,Kost9208G,Moro9209P,Moro1992,Shat9209C,BX9212M,
ACKM9302M,Moro9303I,Miro93122,I1993L,MMS9404U,Moro9502M,AMOS9503A,Kost9509S,
Plef9601S,Akem9606H,BDE0003B,
Seki0212C,Hofm0212S,Laza0303H,KLLR0303C,CT0304Q,CT0307P,CKR0311H,Meht2004R,
Desr0801D,EM0809T,DHS0810Q,Zabr0907R,NS0908Q,IMO0911T}. 

  In this subsection, we give a general discussion of $\beta$-deformed matrix models 
  at finite $N$ (size of matrices) and with generic potentials
 and the attendant noncommutative curve. 
 The curve at the planar level, which the original 
 S-W curve for $SU(N_c)$ gauge group with $2N_c$ flavours are
  relevant to, turn out to come out in a relatively 
 transparent way in the limit. 
     
 Let us begin with the $\beta$-deformed ($\beta$-ensemble of) one-matrix model :  
\be
 Z = \int \de^N \lambda \,
\bigl(\Delta(\lambda)\bigr)^{{+2b_E^2}}
\exp\left(  \frac{b_E}{g_s} \sum_{I=1}^N W(\lambda_I)
\right),
\ee
  where
\be
 \Delta(\lambda) = \prod_{1 \leq I < J \leq N} (\lambda_I - \lambda_J) 
\ee
is the van der monde determinant.

The Virasoro constraints \cite{MM1990O,AJM1990M,Davi1990L,IMat1991N}, 
 namely the Schwinger-Dyson equations of this model for the resolvent,
are obtained by
 inserting $\dis{\sum_{I=1}^{N} \frac{\partial}{\partial \lambda_I} 
 \frac{1}{z - \lambda_I}}$ into $Z$. 
  Adopting the operator notation  of conformal field theory, 
 \begin{align}
J(z) &= \im \partial \phi(z)
= \frac{1}{\sqrt{2} g_s} W'(z) + \sqrt{2} b_E  \, 
\mathrm{Tr} \, \frac{1}{z-M}, \\
T(z) &= - \frac{1}{2} : \partial \phi(z)^2 : + \frac{\im Q_E}{\sqrt{2}}
\partial^2 \phi(z), ~~~ 
Q_E = b_E - \frac{1}{b_E}, 
\end{align}
 they can be written as the vanishing vev of 
 the non-negative part of $T(z)$, 
\be
\label{Tpositive0}
 \text{namely,}~~~{T(z)} \big|_{_{{+}}}, 
 ~~~\langle \! \langle T(z) \big|_{_{{+}}} \rangle \! \rangle = 0. 
\ee
 Eq. \eqref{Tpositive0} can, therefore, be written as
\be
 \langle \! \langle g_s^2 T(z) \rangle \! \rangle 
 = \frac{1}{4} W'(z)^2 - \frac{Q_E}{2} g_s W''(z) -f(z), 
\ee
\be
 f(z) \equiv \left\langle \!\! \left\langle b_E g_s 
 \sum_{I=1}^{N} \frac{W'(z) - W'(\lambda_I)}{z-\lambda_I}
 \right\rangle \!\! \right\rangle. 
\ee

Quite separately, let us introduce the ``curve" $(x,z)=(y(z),z)$ by 
\be
\label{curve}
\left\langle \!\! \left\langle
\left( x + \frac{\im g_s}{\sqrt{2}} \partial \phi(z) \right)
\left( x - \frac{\im g_s}{\sqrt{2}} \partial \phi(z) \right)
\right\rangle \! \! \right \rangle =
x^2 - g_s^2 \langle \! \langle T(z) \rangle \! \rangle =  0.
\ee
 Two remarks are in order.  First of all,  in order for the first equality to be true, $x$ and $z$ must satisfy
 the noncommutative algebra:
\be
[x,z] = Q_E g_s. 
\ee
Second, in order for eq. \eqref{curve} to be algebraic, the singularities in $\langle \! \langle T(z) \rangle \! \rangle$
 must be absent.  This condition is ensured by the Schwinger-Dyson equation
 eq. \eqref{Tpositive0}.

Let us  turn to the $A_{N_c-1}$ quiver matrix model ($\beta$ deformed) which  the effective prepotential  for
 the $SU(N_c)$ gauge theory with $2N_c$ flavours are relevant to.
 This matrix model has been 
constructed \cite{KMMM9208C} such that it  automatically 
 obeys  the $W_{N_c}$ constraints at finite $N_a, \;\; a= 1, \cdots r$, 
 $r=N_c-1$;
\be
 Z \equiv \int \prod_{a=1}^r \left\{ \prod_{I=1}^{N_a} \de \lambda^{(a)}_I \right\}\, 
\left( \Delta_{A_{N_c-1}}(\lambda) \right)^{b_E^2}
\exp\left( \frac{b_E}{g_s}  \sum_{a=1}^r \sum_{I=1}^{N_a} W_a(\lambda^{(a)}_I)
\right), 
\ee
\be
 \Delta_{A_{N_c-1}}(\lambda) = 
 \prod_{a=1}^r \prod_{1 \leq I < J \leq N_a}
 ( \lambda_I^{(a)} - \lambda_J^{(a)})^2
 \prod_{1 \leq a< b \leq r}
 \prod_{I=1}^{N_a} \prod_{J=1}^{N_b}
 ( \lambda^{(a)}_I - \lambda^{(b)}_J )^{(\alpha_a, \alpha_b)} . 
\ee
 We follow the logic of $\beta$-deformed one-matrix model at finite $N_a$.
  In this model, there exists $N_c$ spin 1 currents  that satisfy 
  $\dis{\sum_{i=1}^{N_c} J_i(z) = 0}$: 
\begin{align}
 &J_i(z) = \im \partial \varphi_i (z) = \frac{1}{g_s} t_i(z) 
 + b_E \sum_{a=1}^{N_c-1}
 {( \delta_{i,a} - \delta_{i,a+1} )} \mathrm{Tr} \, \frac{1}{z - M_a }, \\
 &t_i(z) = \sum_{a=i}^{N_c-1} W'_a(z) 
 - \frac{1}{n} \sum_{a=1}^{N_c-1}
 a \, W_a'(z). 
\end{align}
Note that
\be
 \dis{:\det( x - \im g_s \partial \phi(z)):
 = : \prod_{1 \leq i < N_c}^{\leftarrow} ( x - g_s J_i(z) ) :}
\ee 
contains $W_{N_c}$ generators  and
 the $W_{N_c}$ constraints are expressible as
\be
 \left\langle \!\! \left\langle
 \det( x - \im g_s \partial \phi(z)) \big|_{_{+}}
 \right\rangle \!\! \right\rangle = 0. 
\ee
The curve $\Sigma$ $(x=y_i(z), z)$  that we postulate 
 in \cite{IMO0911T} is
\be
 {\big\langle \!\! \big\langle \det( x - \im g_s \partial \phi(z)) 
 \big\rangle \!\! \big\rangle =0}. 
\ee
The isomorphism with the Witten-Gaiotto curve has been established
 by taking the planar limit of this construction 
  as we will see in the next subsection. 
In fact, the planar limit implies the singlet factorization 
 which assigns  the $c$ number value
 to the operator $\partial \phi(z)$ and the curve factorizes as  
\be
 {0 = \prod_{i=1}^{N_c} (x - y_i(z))} \sim {\rm lim} g_s J_i, ~~~~~~~
 (x,z) = (y_i(z),z). 
\ee
where 
\beq \label{yiz}
y_i(z) := \lim_{g_s \rightarrow 0}
\im g_s \langle \! \langle \partial \varphi_i(z) \rangle \! \rangle.  
\eeq

\subsection{the three Penner potential and the agreement with the Witten-Gaiotto curve}
The list of papers which discuss subjects closely related to 
 that of this subsection include 
\cite{Penn1987T,Penn1988P,Witt9703S,MMM9706I,
Gaio0904N,GM0904T,Tach0905S,BBT0906W,AGT0906L,NX0907N,
Wyll0907A,DMO0907L,MTTY0907N,MMM0907C,GMN0907W,
Gaio0908A,MMMM0908C,MM0908T,MM0908O,NX0908O,AGGT0909L,
DGOT0909G,BTW0909S,MMM0909O,DV0909T,MMM0909Z,Pogh0909R,
MM0909P,BT0909H,AY0910F,Papa0910T,MM0910N,IMO0911T,AM0911N,Gaio0911S,
NX0911H,MM0911N,ABMM1410S,GMM1410W}. 

Let us specialize our discussion to the three Penner model.  Choose the potential as
\be
 W_a (z) = \sum_{p=1}^{3} (\mu_p, \alpha_a) \log (q_p - z),~~~~~~~~  
 {q_0 = \infty,~  q_1 = 0,~  q_2=1,~  q_3=q}. 
\ee
The matrix integrals of this case realize the integral representation of the conformal block and
the size of each matrix corresponds with the number 
 of screening charges we have to insert to built the block. 
  As is clear from the discussion above,
the planar spectral curve of the $A_{N_c-1}$ quiver matrix model takes the form
\be \label{curve:MM}
 x^{N_c} 
 = \sum_{k=2}^{N_c} \frac{(-1)^{k-1} Q_k(z)}{(z(z-1)(z-q))^k} x^{N_c-k},
\ee
for some polynomials $Q_k(z)$ in $z$.

On the other hand, 
the Seiberg-Witten curve  for the case of $SU(N_c)$ gauge theory with $2N_c$ 
 massive flavour multiplets, originally proposed in \cite{Witt9703S}, 
 can get converted into the Gaiotto form \cite{Gaio0904N} by 
\begin{align} \label{curve:SW}
x^{N_c} 
&= \sum_{k=2}^{N_c} \frac{P_{2k}^{(k)}(t)}{( t(t-1)(t-q_{\mathrm{bare}}))^k}
x^{N_c-k},
\end{align}
where $P_{2k}^{(k)}(t)$ are degree $2k$ polynomials in $t$. 
The two curves eq. \eqref{curve:MM} and eq. \eqref{curve:SW}  
 are evidently similar. 
We can also see that the residues of  
 $y_i(z) \de z ~(i=1, \cdots, N_c)$ at $z=1,~q,~0,~\infty$ and 
 those of $x \de t$ at $t=1,~q_{\rm bare},~0,~\infty$ on the $i$-th sheet
 can be equated.

For general $N_c$, these residues in fact
  match if the weights of the vertex operators are
  identified with the mass parameters of the gauge theory by the following relations
  \cite{IMO0911T} : 
\begin{align}
\mu_0 = \sum_{a=1}^{N_c-1} ( - m_a + m_{a+1}) \Lambda^a, \qquad
\mu_1 = \sum_{a=1}^{N_c-1} ( \widetilde{m}_a - \widetilde{m}_{a+1}) \Lambda^a,\\
\mu_2 = \left( \sum_{i=1}^{N_c} m_i \right) \Lambda^1, \qquad
\mu_3 = \left( \sum_{i=1}^{N_c} \widetilde{m}_i \right) \Lambda^{N_c-1}.
\end{align}
The matrix model potentials $W_a(z)~ (a=1,2,\dots,N_c-1)$ are fixed as
\begin{align}
 W_a(z) = ( \widetilde{m}_a - \widetilde{m}_{a+1})
 \log z + \delta_{a,1} \left( \sum_{i=1}^{N_c} m_i \right) \log (1-z) \notag \\ 
 + \delta_{a,N-1} \left( \sum_{i=1}^{N_c} \widetilde{m}_i \right) 
 \log ( q_{\mathrm{UV}} - z).
\end{align}
With this choice of the multi-log potentials, the $A_{N_c-1}$ quiver matrix model curve 
 in the planar limit coincides with the $SU(N_c)$ Seiberg-Witten curve 
  with $2N_c$ massive hypermultiplets.

\subsection{direct evaluation of the matrix integral as Selberg integral}

The list of papers which discuss subjects closely related to 
 that of this subsection include 
\cite{Selb1944R,Suth1971E,Suth1972E,DF1984C,BPZ1984I,
Macd1987C,Stan1989S,Kane1993S,AMOS9411C,AMOS9503E,MY1995S,
Kade1997T,Macd1998S,LMN0302S,SSAF0407C,KM0712I,
KMST0911N,EM0911P,MMS0911M,SW0911A,Giri0912O,AM0912C,FHT0912G,
Taki0912O,Sulk0912M,Shak0912E,Popo1001O,IO51003M,
MMM1003C,KPW1004A,HJS1004P,MS1004T,
NX1005H,Tai1006T,
EM1006S,NX1006N,KMS1007A,Taki1007S,IOY1008M,MY1009S,BMS1010T,
MMS1010O,CDV1010N,MMS1011B,MMM1011O,MMS1011T,BMTY1011G,
Marc1012A,Mars1101O,BB1102A,Piat1102C,MMPS1103R,IY1104e,BMT1104Q,
IO1106A,SWY1107A,EPSS1110C,
Shib1111N,NR1112b,BPTY1112M,Moro1201C,Moro1204F,Bour1206L,
NR1207M,KMZ1207V,Kref1209P,Bour1212L,FP1212P,KMST1301W,
Baek1303G,Nemk1307S,
Bour1310N,Bour1402C,AM1403T,
Smir1404P,
IMM1406M,Maru1412b,RZ1504C,RZ1506C}. 

In this subsection, we consider 2-d conformal field theory
 which has the Virasoro symmetry with the central charge $c$. 
The correlation functions for primary operators 
 $\Phi_{\Delta}(z, \bar{z})$
 with the conformal weight $\Delta$ are strongly constrained by this symmetry.
We are interested in the four-point functions 
 which can be expressed as 
\beq 
\langle \Phi_{\Delta_1}(\infty, \infty) \Phi_{\Delta_2}(1,1)
 \Phi_{\Delta_3}(q, \bar{q}) \Phi_{\Delta_4}(0,0) \rangle 
 = \sum_I C_{\Delta_1 \Delta_2}^{\Delta_I} K_{\Delta_I} C_{\Delta_3 \Delta_4}^{\Delta_I}
 \Bigl| {\mathcal{F}}(q|c; \Delta_1, \Delta_2, \Delta_3, \Delta_4, \Delta_I) \Bigr|^2. 
\eeq
The sum on $I$ is taken over all possible internal states. 
Here $K_{\Delta}$ and $C_{\Delta_1\Delta_2}^{\Delta_3}$ 
 are the model-dependent factors.
In contrast, the conformal block \footnote{For a review, \cite{Alva1991S,ZZ2009C}.} 
 denoted by 
 $\CF (q|c; \Delta_1, \Delta_2, \Delta_3, \Delta_4, \Delta_I)$ 
 is a model-independent and
 purely representation theoretic quantity,   
\beq \label{4ptCB}
 \mathcal{F}(q|c; \Delta_1, \Delta_2, \Delta_3, \Delta_4, \Delta_I) 
= \sum_{|Y|=|Y'|} q^{|Y|}
\gamma_{\Delta_I, \Delta_1, \Delta_2}(Y)
Q_{\Delta_I}^{-1}(Y,Y') \gamma_{\Delta_I, \Delta_3, \Delta_4}(Y'), 
\eeq 
where $Q_{\Delta}(Y,Y') = \langle \Delta | L_Y L_{-Y'} | \Delta \rangle$
 is the Shapovalov form with $L_{Y} = L_{k_1} L_{k_2} \dotsm L_{k_{\ell}}$
 for partition $Y = (k_1, k_2, \dots, k_{\ell})$ and 
\beq 
 \gamma_{\Delta, \Delta_1, \Delta_2}(Y)
=  \prod_{i=1}^{\ell} \left( 
 \Delta + k_i \Delta_1 - \Delta_2 + \sum_{j < i } k_j \right). 
\eeq
 
Let us consider the four-point conformal block on sphere,  
\be
 \mathcal{F}(q|c;\Delta_1,\Delta_2,\Delta_3,\Delta_4,\Delta_I),  
 \label{CB}
\ee
with
\be 
 c=1-6Q_E^2, ~~ \Delta_i = \frac{1}{4}\alpha_i(\alpha_i-2Q_E), 
 ~~ \Delta_I=\frac{1}{4}\alpha_I(\alpha_I-Q_E).  
\ee
The parameter $\alpha_4$ is determined by the following
 momentum conservation condition which comes from the zero-mode part:  
\be \label{mom-con}
 \alpha_1 + \alpha_2 + \alpha_3 + \alpha_4 + 2 (N_L + N_R) b_E = 2 Q_E. 
\ee 
The internal momentum $\alpha_I$ is given by  
\be 
 \alpha_I = \alpha_1 + \alpha_2 + 2 N_L b_E    
 = - \alpha_3 - \alpha_4 - 2 N_R b_E + 2 Q_E.
\ee
Eq. \eqref{CB} has an integral representation as a version of
 $\beta$-deformed matrix model.  
Actually, the Dotsenko-Fateev multiple integrals,     
\begin{align} \label{pert-selb}
 Z_{\text{pert-}(\mathrm{Selberg})^2}&
  (q\, | \,  b_E; N_L, \alpha_1, \alpha_2; N_R, \alpha_4, \alpha_3) 
  = q^{\Delta_I-\Delta_1-\Delta_2} (1 - q)^{(1/2)\alpha_2 \alpha_3} \cr
& \times \left(  \prod_{I=1}^{N_L} \int_0^1 \de x_I \right)
\prod_{I=1}^{N_L} x_I^{b_E \alpha_1} ( 1 - x_I)^{b_E \alpha_2}
(1 - q x_I )^{b_E \alpha_3}
\prod_{1 \leq I < J \leq N_L}
| x_I - x_J |^{2b_E^2} \cr
& \times
\left( \prod_{J=1}^{N_R} \int_0^1 \de y_J \right)
\prod_{J=1}^{N_R} y_J^{b_E \alpha_4}
(1 - y_J )^{b_E \alpha_3}
( 1 - q \, y_J )^{b_E \alpha_2}
\prod_{1 \leq I < J \leq N_R}
| y_I - y_J|^{2b_E^2} \cr
& \qq \qq \qq \qq \times
\prod_{I=1}^{N_L} \prod_{J=1}^{N_R} ( 1 - q\, x_I y_J )^{2b_E^2}, 
\end{align} 
are regarded as a free field representation of eq. \eqref{CB}. 

From now on, we follow the discussion of \cite{IO51003M}. 
Eq. \eqref{pert-selb} is in fact partition function of the 
 ``perturbed double-Selberg matrix model". 
If we forget the Veneziano factor 
$q^{\Delta_I-\Delta_1-\Delta_2}(1-q)^{(1/2)\alpha_2\alpha_3}$,
we  see that at $q=0$ this expression decouples into two independent
Selberg integrals. 
In order to develop its $q$-expansion, 
 it is more convenient to interpret this
 multiple integrals as perturbation of the products of the two Selberg integrals.

We have the following expression of the perturbed double-Selbarg model: 
\begin{align} \label{pert-selb:BB}
& Z_{\text{pert-}(\mathrm{Selberg})^2} 
(q\, |\, b_E; N_L, \alpha_1, \alpha_2; N_R, \alpha_4, \alpha_3) \cr
&~~~~~ = q^{\Delta_I - \Delta_1 - \Delta_2}\,
\mathcal{B}_0(b_E; N_L, \alpha_1, \alpha_2; N_R, \alpha_4, \alpha_3)\,
\mathcal{B}(q\, |\, b_E; N_L, \alpha_1, \alpha_2; N_R, \alpha_4, \alpha_3), 
\end{align}
where 
\begin{align}
&\mathcal{B}_0(b_E; N_L, \alpha_1, \alpha_2; N_R, \alpha_4, \alpha_3) \\
&~~~~ = S_{N_L}(1 + b_E \alpha_1, 1 + b_E \alpha_2, b_E^2)\,
S_{N_R}(1 + b_E \alpha_4, 1 + b_E \alpha_3, b_E^2), \\
& \mathcal{B}(q\, |\,  b_E; N_L, \alpha_1, \alpha_2; N_R, \alpha_4, \alpha_3) \cr
&~~~~ = (1-q)^{(1/2)\alpha_2 \alpha_3}
{\bigg\langle \!\!\! \bigg\langle}
\prod_{I=1}^{N_L} (1 - q x_I)^{b_E \alpha_3}
\prod_{J=1}^{N_R} ( 1 - q y_J)^{b_E \alpha_2}
\prod_{I=1}^{N_L}
\prod_{J=1}^{N_R} ( 1 - q x_I y_J)^{2b_E^2}
{\bigg\rangle\!\!\!\bigg\rangle_{\!\!{N_L, N_R}}}. 
\label{CBres}
\end{align}
Here $S_{N_L}$ and $S_{N_R}$ are the celebrated Selberg integral  
\be \label{SelI}
\begin{split}
& S_N(\beta_1, \beta_2, \gamma)
= \left( \prod_{I=1}^N \int_0^1 \de x_I \right)
\prod_{I=1}^{N} x_I^{\beta_1 -1} (1-x_I)^{\beta_2-1}
 \prod_{1 \leq I < J \leq N} | x_I - x_J |^{2\gamma},
\end{split}
\ee
and the averaging $\langle\!\!\!\langle \cdots \rangle\!\!\!\rangle_{N_L,N_R}$
 is taken with respect to the unperturbed Selberg matrix model, 
\be 
\begin{split}
& Z_{(\mathrm{Selberg})^2}(b_E; N_L, \alpha_1, \alpha_2; N_R,
\alpha_4, \alpha_3) \cr
&= Z_{\mathrm{Selberg}}(b_E; N_L, \alpha_1, \alpha_2)
Z_{\mathrm{Selberg}}(b_E; N_R, \alpha_4, \alpha_3) \cr
&:= S_{N_L}(1 + b_E \alpha_1, 1 + b_E \alpha_2, b_E^2)\,
S_{N_R}(1 + b_E \alpha_4, 1 + b_E \alpha_3, b_E^2).
\end{split}
\ee
Below we also use   
$\langle\!\!\!\langle \cdots \rangle\!\!\!\rangle_{N_L}$ and 
$\langle\!\!\!\langle \cdots \rangle\!\!\!\rangle_{N_R}$ 
which imply the averaging with respect to $Z_{\mathrm{Selberg}}(N_L)$
 and to $Z_{\mathrm{Selberg}}(N_R)$, respectively. 

The function 
 $\mathcal{B}(q) = \mathcal{B}(q\, |\, b_E; N_L, \alpha_1, \alpha_2;N_R, \alpha_4, \alpha_3)$ 
 has the following $q$-expansion \cite{IO51003M}:  
\begin{align}
{\mathcal{B}(q)} &= 1 + \sum_{\ell=1}^{\infty} q^{\ell}\, \mathcal{B}_{\ell} \cr
&{\ = \left\langle\!\!\!\left\langle
\exp\left[ - 2 \sum_{k=1}^{\infty}
\frac{q^k}{k}
\left( b_E \sum_{I=1}^{N_L} x_I^k + \frac{1}{2} \alpha_2 \right)
\left( b_E \sum_{J=1}^{N_L} y_J^k + \frac{1}{2} \alpha_3 \right)
\right]
\right\rangle\!\!\!\right\rangle_{\!\! N_L, N_R}} \\
&= (1-q)^{(1/2)\alpha_2 \alpha_3} \mathcal{A}(q), 
\end{align}
where we have defined ${\mathcal{A}(q)}$ by 
\begin{align}
{\mathcal{A}(q)} &= 1 + \sum_{\ell=1}^{\infty} q^{\ell}\, \mathcal{A}_{\ell} \cr
&={ \left\langle\!\!\!\left\langle
\exp
\left[
- \sum_{k=1}^{\infty} \frac{q^k}{k}
\left( \alpha_2 + b_E \sum_{I=1}^{N_L} x_I^k \right)
\left( b_E \sum_{J=1}^{N_R} y_J^k \right) 
\right. \right. \right.} \cr
& \qq \qq {\left. \left. \left.
- \sum_{k=1}^{\infty} \frac{q^k}{k}
\left( b_E \sum_{I=1}^{N_L} x_I^k \right)
\left( \alpha_3 + b_E \sum_{J=1}^{N_R} y_J^k \right)
\right]
\right\rangle\!\!\!\right\rangle_{\!\! N_L, N_R}}. 
\end{align}
It takes form  
\be 
 \mathcal{A}(q) = \sum_{k=0}^{\infty} q^k
\sum_{|Y_1|+ |Y_2|=k} \mathcal{A}_{Y_1, Y_2}. 
\ee
Note that a pair of partitions $(Y_1, Y_2)$ naturally appears.  

In general, the following correlation function is 
 calculable:  
\be \label{ZpS}
Z_{\mathrm{pert}-\mathrm{Selberg}}(\beta_1, \beta_2, \gamma; \{ g_i \}  )
:=S_N(\beta_1, \beta_2, \gamma)
\left\langle\!\!\!\left\langle
\exp\left( \sum_{I=1}^{N} W(x_I; g )\right)
\right\rangle\!\!\!\right\rangle_{N},
\ee
with 
\be
 W(x; g )= \sum_{i=0}^{\infty} g_i x^i.
\ee 
The averaging is with respect to the Selberg integral eq. \eqref{SelI}.   
The exponential of the potential is expanded 
 by the Jack polynomial
\be
 \exp\left( \sum_{I=1}^{N} W(x_I; \{ g_i \} )\right)
 = \sum_{\lambda} C_{\lambda}^{(\gamma)}(g) \,
 P_{\lambda}^{(1/\gamma)}(x), 
\ee
where $P_{\lambda}^{(1/\gamma)}(x)$ is a polynomial
of $x=(x_1, \dotsm, x_N)$ and  
$\lambda=(\lambda_1, \lambda_2, \dotsm)$ is a partition:
$\lambda_1 \geq \lambda_2 \geq \dotsm \geq 0$. 
The Jack polynomial is characterized as the eigenstates of
\be
\sum_{I=1}^{N}
\left( x_I \frac{\partial}{\partial x_I} \right)^2
+ \gamma \sum_{1 \leq I <  J \leq N}
\left( \frac{x_I + x_J}{x_I - x_J} \right)
\left(
x_I \frac{\partial}{\partial x_I}
- x_J \frac{\partial}{\partial x_J}
\right),
\ee
with homogeneous degree $|\lambda|=\lambda_1+\lambda_2+
\dotsm$
and is normalized such that 
\be
P_{\lambda}^{(1/\gamma)}(x) = m_{\lambda}(x) + \sum_{\mu < \lambda}
a_{\lambda \mu} m_{\mu}(x).
\ee
Here $\mu < \lambda$ stands for the dominance ordering
 defined by 
\beq
 |\mu| = |\lambda| ~~~~~ \text{and} ~~~~~ 
 \mu_1 + \mu_2 + \cdots + \mu_n 
 < \lambda_1 + \lambda_2 + \cdots + \lambda_n 
 ~~~ \text{for all $n \geq 1$,}
\eeq
and 
$m_{\lambda}(x)$ is the monomial symmetric function. 
Explicit forms of the Jack polynomials for $|\lambda| < 2$ are as follows:
\begin{align} \label{Jack}
P_{(1)}^{(1/\gamma)}(x) &= m_{(1)}(x) = \sum_{I=1}^{N} x_I, \cr
P_{(2)}^{(1/\gamma)}(x) &= m_{(2)}(x)
+ \frac{2\gamma}{1+\gamma}\, m_{(1^2)}(x)
= \sum_{I=1}^{N} x_I^2 + \frac{2\gamma}{1+\gamma} \sum_{1\leq I <J \leq N}
x_I x_J, \cr
P_{(1^2)}^{(1/\gamma)}(x)&= m_{(1^2)}(x) =
\sum_{1\leq I <J \leq N}
x_I x_J.
\end{align}
The Selberg average for single Jack polynomial is known as 
 Macdonald-Kadell integral \cite{Macd1987C,Kade1997T,Kane1993S} 
 which implies that
\be \label{MK}
\begin{split}
\Bigl\langle\!\!\Bigr\langle P_{\lambda}^{(1/\gamma)}(x)
\Bigr\rangle\!\! \Bigr\rangle_{N}
&= \prod_{i \geq 1}
\frac{\dis
\bigl(\, \beta_1+ (N-i) \gamma \, \bigr)_{\lambda_i} \
\bigl(\, (N + 1-i) \gamma \, \bigr)_{\lambda_i}}
{\dis
\bigl(\, \beta_1 + \beta_2 + (2 N-1 - i) \gamma \,
\bigr)_{\lambda_i}} \cr
& \times
\prod_{(i,j) \in \lambda}
\frac{1}{(\lambda_i - j + ( \lambda_j' -i +1 ) \gamma)},
\end{split}
\ee
where $(a)_n$ is the Pochhammer symbol:
\be
(a)_n = a (a+1) \dotsm (a+n-1), \qq
(a)_0 = 1, 
\ee
and $\lambda'$ stands for the conjugate partition of $\lambda$. 

In order to apply this to eq. \eqref{pert-selb:BB},
 let us set $\gamma=b_E^2$ and
\be
N \rightarrow N_L, \qq
\beta_1 \rightarrow 1+b_E \alpha_1, \qq
\beta_2 \rightarrow 1+b_E \alpha_2,
\ee
for the ``left'' part.
Similar replacement yields the expression for the ``right" part.  
We obtain   
\begin{align}
\Bigl\langle\!\!\Bigr\langle P_{\lambda}^{(1/b_E^2)}(x)
\Bigr\rangle\!\! \Bigr\rangle_{N_L}
&= \prod_{i \geq 1}
\frac{\dis
\bigl(\, 1 + b_E \alpha_1 + b_E^2 (N-i)  \, \bigr)_{\lambda_i} \
\bigl(\, b_E^2(N_L + 1-i)\, \bigr)_{\lambda_i}}
{\dis
\bigl(\, 2 + b_E(\alpha_1 + \alpha_2) + b_E^2 (2 N_L-1 - i) \,
\bigr)_{\lambda_i}} \cr
& \times
\prod_{(i,j) \in \lambda}
\frac{1}{(\lambda_i - j + b_E^2 ( \lambda_j' -i +1 ))}. 
\end{align}
From the explicit form of Jack polynomials 
 for $|\lambda|<2$ listed in eq. \eqref{Jack}, 
we obtain \cite{IO51003M}
\begin{align}
\left\langle\!\!\!\left\langle
b_E \sum_{I=1}^{N_L} x_I
\right\rangle\!\!\!\right\rangle_{\!\! N_L}
&= \frac{b_E N_L( b_E N_L - Q_E + \alpha_1)}
{(\alpha_I - 2 Q_E)}, \label{Jack1:ave}\\
2 \left\langle\!\!\!\left\langle
b_E^2 \!\!\! \sum_{1 \leq I < J \leq N_L} \!\!\! x_I x_J
\right\rangle\!\!\!\right\rangle_{N_L} 
&= \frac{b_E N_L ( b_E N_L - b_E)(\alpha_1 + b_E N_L -Q_E)
(\alpha_1 + b_E N_L - Q_E - b_E)}
{(\alpha_I - 2 Q_E)(\alpha_I - 2 Q_E - b_E)}, \\
\left\langle\!\!\!\left\langle
b_E \sum_{I=1}^{N_L} x_I ( 1 - x_I )
\right\rangle\!\!\!\right\rangle_{N_L} 
&= \frac{b_E N_L(\alpha_1+ b_E N_L - Q_E)(\alpha_2 + b_E N_L - Q_E)
( \alpha_1 + \alpha_2 + b_E N_L - 2 Q_E)}
{(\alpha_I - 2 Q_E)(\alpha_I - 3 Q_E+b_E)(\alpha_I - 2 Q_E - b_E)}. 
\end{align}

Recall, at $q = 0$, the perturbed double-Selberg matrix model 
 reduces to a pair of decoupled Selberg integrals.   
The original model ($q \neq 0$) is built through the resolvents 
 as in eq. \eqref{CBres}. 
For definiteness, let us consider the left-part,  
\begin{align}
Z_{\mathrm{Selberg}}
(b_E; N_L, \alpha_1, \alpha_2)
&= \left( \prod_{I=1}^{N_L} \int_0^1 \de x_I \right)
\prod_{1 \leq I < J \leq N_L} | x_I - x_J|^{2b_E^2} 
\exp\left( b_E \sum_{I=1}^{N_L} \widetilde{W}(x_I) \right). 
\end{align}
where 
\begin{align} 
\widetilde{W}(x) &= \alpha_1 \log x + \alpha_2 \log (1-x).  
\end{align}
By inserting 
\be 
 \sum_{I=1}^{N_L} \frac{\partial}{\partial x_I} \frac{1}{z - x_I}, 
\ee
into the integrand, we obtain the loop equation at finite $N$, 
\be \label{LoopEq}
 \left\langle \!\!\! \left\langle \Bigl( \widehat{w}_{N_L}(z) \Bigr)^2
 \right\rangle\!\!\!\right\rangle_{\!\! N_L}
 + \left( \widetilde{W}'(z) + Q_E \frac{\de }{\de z} \right)
 \left\langle \!\! \left\langle \Bigl. \widehat{w}_{N_L}(z) \Bigr.
 \right\rangle\!\! \right\rangle_{\!\! N_L} - \tilde{f}_{N_L}(z) = 0, 
\ee 
where
\be 
 \widehat{w}_{N_L}(z):= b_E \sum_{I=1}^{N_L} \frac{1}{z- x_I},
 ~
 \tilde{f}_{N_L}(z):=
 \left\langle\!\!\!\left\langle b_E \sum_{I=1}^{N_L}
 \frac{ \widetilde{W}'(z) - \widetilde{W}'(x_I)}
 {z - x_I }
 \right\rangle\!\!\!\right\rangle_{\!\! N_L}. 
\ee
The expectation value of $\widehat{w}_{N_L}(z)$ is 
 the finite $N$ resolvent 
\be 
 \widetilde{w}_{N_L}(z) := \left\langle \!\! \left\langle \Bigl. 
 \widehat{w}_{N_L}(z)
 \right\rangle\!\!\right\rangle_{\!\! N_L}
 = \left\langle \!\!\! \left\langle b_E
 \sum_{I=1}^{N_L} \frac{1}{z - x_I}
 \right\rangle\!\!\!\right\rangle_{\!\! N_L}. 
\ee
By looking at $O(1/z), O(1/z^2), O(1/z^3)$, 
 we obtain the exact results:  
\begin{align}
\bigl\langle\!\bigl\langle b_E \, p_{(1)}(\mu)
\bigr\rangle \!\bigr\rangle_{N_L}
=
\left\langle\!\!\!\left\langle
b_E \sum_{I=1}^{N_L} x_I
\right\rangle\!\!\!\right\rangle_{\!\! N_L}
&= \frac{b_E N_L( b_E N_L - Q_E + \alpha_1)}{(\alpha_1 + \alpha_2 + 2 b_E N_L - 2 Q_E)}, \cr
\tilde{f}_{N_L}(z) &= - \frac{b_E N_L(\alpha_1+\alpha_2 + b_E N_L - Q_E)}{z(z-1)}, \cr
- \widetilde{w}_{N_L}(0)=
\left\langle\!\!\!\left\langle
b_E \sum_{I=1}^{N_L} \frac{1}{x_I}
\right\rangle\!\!\!\right\rangle_{\!\! N_L}
&=\frac{b_E N_L(\alpha_1+\alpha_2 + b_E N_L - Q_E)}{\alpha_1}, \cr
\widetilde{w}_{N_L}(1)=
\left\langle\!\!\!\left\langle
b_E \sum_{I=1}^{N_L} \frac{1}{1 - x_I}
\right\rangle\!\!\!\right\rangle_{\!\! N_L}
&=\frac{b_E N_L(\alpha_1+\alpha_2 + b_E N_L - Q_E)}{\alpha_2}. 
\end{align}
The first one agrees with eq. \eqref{Jack1:ave}.

Now, let us determine the 0d-4d dictionary. 
In the matrix model (0d side), 
 we have seven parameters with one constraint eq. \eqref{mom-con}:    
\be 
 b_E,~ N_L,~ \alpha_1,~ \alpha_2,~ N_R,~ \alpha_4,~ \alpha_3, 
\ee  
while in $\CN=2,~ SU(2), N_f=4$ gauge theory (4d side), 
 there exists six unconstrained parameters:  
\be 
 \frac{\epsilon_1}{g_s}, \ \ 
 \frac{a}{g_s}, \ \ 
 \frac{m_1}{g_s}, \ \ 
 \frac{m_2}{g_s}, \ \ 
 \frac{m_3}{g_s}, \ \
 \frac{m_4}{g_s}. 
\ee 
Here $a$ is the vacuum expectation value of the adjoint scalar,
$m_i$ are mass parameters and $\epsilon_1$ is one of the Nekrasov's
deformation parameter. 
By looking at $\mathcal{B}_1 = \mathcal{A}_1 - \frac{1}{2} \alpha_2 \alpha_3$
 and the explicit form of $\mathcal{A}_1^{\mathrm{Nek}} = 
\mathcal{A}_{[1],[0]}^{\mathrm{Nek}} + 
\mathcal{A}_{[0],[1]}^{\mathrm{Nek}}$, 
\begin{align}
 \mathcal{A}_{[1],[0]}^{\mathrm{Nek}}
 &= \frac{(a+m_1)(a+m_2)(a+m_3)(a+m_4)}{2a(2a+\epsilon) g_s^2}, \\
 \mathcal{A}_{[0],[1]}^{\mathrm{Nek}}
 &= \frac{(a-m_1)(a-m_2)(a-m_3)(a-m_4)}{2a(2a-\epsilon) g_s^2}, 
\end{align}
we obtain  
\begin{align} 
b_E N_L &= \frac{a-m_2}{g_s},&
b_E N_R &= - \frac{a+m_3}{g_s}, \cr 
\alpha_1 &= \frac{1}{g_s}( m_2 - m_1 + \epsilon ),&
\alpha_2 &= \frac{1}{g_s}( m_2 + m_1 ), \cr
\alpha_3 &= \frac{1}{g_s}( m_3
 + m_4 ),&
\alpha_4 &= \frac{1}{g_s}( m_3 - m_4 + \epsilon ).
\end{align}
The first two formulas tell us clearly the necessity that 
 the filling fractions of the $\beta$-deformed matrix model 
 must be explicitly specified at finite $N$ 
 in order to exhibit the Coulomb moduli. 

In the next order, the expansion coefficients $\CA_2$ are rearranged as
\be 
 \mathcal{A}_2 = \sum_{|Y_1|+|Y_2|=2} \mathcal{A}_{Y_1,Y_2}=
 \mathcal{A}_{(2),(0)} 
 + \mathcal{A}_{(1^2), (0)}
 + \mathcal{A}_{(1),(1)} 
 + \mathcal{A}_{(0),(1^2)} 
 + \mathcal{A}_{(0),(2)},
\ee
where
\be 
 \mathcal{A}_{Y_1, Y_2}
 = \Bigl\langle\!\! \Bigl\langle
 \ M_{Y_1,Y_2}(x) \ 
 \Bigr\rangle\! \! \Bigr\rangle_{\! N_L}
 \Bigl\langle\!\! \Bigl\langle
 \ \widetilde{M}_{Y_1,Y_2}(y) \ 
 \Bigr\rangle\! \! \Bigr\rangle_{\! N_R}. 
\ee
Unfortunately, finding $M$ and $\widetilde{M}$ are not straightforward. 
But at least for $|Y_1| + |Y_2| \leq 2$, 
 the explicit forms for them have been obtained. For examples, 
\be 
 \widetilde{M}_{(2),(0)}(y) =  b_E^2 P_{(2)}^{(1/b_E^2)}(y),~~~~~~~~ 
 \widetilde{M}_{(1^2),(0)}(y)= \frac{2 b_E^2}{1+b_E^2}
 P_{(1^2)}^{(1/b_E^2)}(y). 
\ee

\begin{figure}[H]
\centering
  \includegraphics[height=7cm]{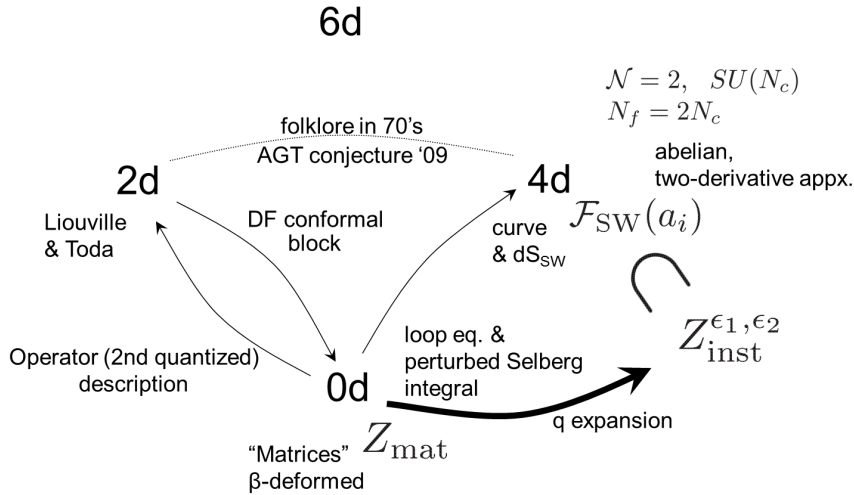}
\caption{Bird view of the 2d-4d connection through matrices 
\label{0d2d4d}
}
\end{figure}

We illustrate our discussion in this section by Fig. \ref{0d2d4d}. 

\subsection{more recent developments}
The list of papers which discuss subjects closely related to 
 that of this subsection include 
\cite{FZ1985P,FZ1987R,GQ1987M,KN1990Y,
SKAO9507A,FF9508Q,AKOS9508Q,AKMO9604V,FR9706T,
FMP0406M,AK0805R,AY0910F,FL0912O,HJS1004P,AY1004F,BFL1011P,MMS1012A,MMSS1105P,
BF1105S,Kimu1105M,NT1106C,BMT1106I,BBB1106I,BMT1107G,Wyll1109C,
Ito1110R,ZM1110S,AT1110P,BBFL1111I,SV1202C,BW1205N,
FMP1210D,BBT1211B,ABT1306C,KMZ1306E,MS1307F,IOY213082,NPPT13125,MMZ1312G,
Ohku1404E,MRZ1405C,IOY31408q,Spod1409A,MP1409T,MPTY1411F,
IMP1412T,Zenk1412G,Szab1507N}. 

We have reviewed the 2d-4d connection from a view point of the matrix model. 
In this subsection, 
 we comment on some of the more recent developments. 

In the last subsection, we have presented the connection between 
 the Virasoro conformal blocks and the four-dimensional 
 $SU(2)$ instanton partition functions 
 via the matrix model and the Selberg integral. 
This discussion has been generalized in part to that  
 between the $W_{N}$ blocks and the $SU(N)$ partition functions \cite{ZM1110S}.

The both sides also have a natural generalization 
 as a $q$-lift \cite{AY1004F}. 
The Virasoro/$W_{N}$ symmetry in the two-dimensional CFT side   
 is deformed to the $q$-deformed Virasoro/$W_{N}$ symmetry 
 while the four-dimensional $SU(N)$ gauge theory
  is lifted to the five-dimensional theory.  
It is interesting to consider 
 the root of unity limit $q \to e^{\frac{2\pi \im}{r}}$ 
 of the $q$-Virasoro/W$_{N}$ algebras. 
The appropriate limiting procedure  \cite{IOY213082,IOY31408q}
 to the root of unity exhibits the connection 
 between the super Virasoro ($r=2$) or the $\textbf{Z}_r$-parafermionic CFT 
 and the gauge theory  on $\textbf{R}^4/\textbf{Z}_r$ \cite{BF1105S,NT1106C}. 

There are several pieces of work \cite{FL0912O,HJS1004P,SV1202C,KMZ1306E} 
 which prove the 2d-4d connection. 
The explicit identification can be established in the case of $\beta = 1$
\cite{MMS1012A,MMSS1105P}. 
In order to apply to the $\beta \neq 1$ case, 
 the conformal blocks have to be expanded by the generalized Jack polynomial 
  \cite{MS1307F} that modifies the standard one. 
For some lower rank cases, this has been explicitly constructed \cite{MMZ1312G}.

\section*{Acknowledgment}
We gratefully acknowledge the valuable discussion with 
 S. Aoyama, H. Awata, K. Fujiwara, H. Kanno, H. Kawai, A. Marshakov, N. Maru,
 K. Maruyoshi, M. Matone, Y. Matsuo, A. Mironov, A. Morozov, T. Nakatsu, K. Ohta, 
 T. Oota, M. Sakaguchi, S. Seki, D. Serban, M. Taki, Y. Yamada and N. Yonezawa.

\end{document}